\documentclass[pra,showpacs,groupedaddress,amssymb,twocolumn,notitlepage,nofootinbib,longbibliography,floatfix,superscriptaddress]{revtex4-1}
\usepackage{graphicx,amsmath,dsfont}
\usepackage[dvipsnames]{xcolor}
\usepackage{subcaption}
\usepackage{comment}
\usepackage{enumerate}
\usepackage{float}
\usepackage{hyperref}
\hypersetup{
    colorlinks=true,
    linkcolor=red,
    filecolor=magenta,      
    urlcolor=blue,
    pdftitle={Decoherence Limits the Cost to Simulate an Anharmonic Oscillator},
    pdfpagemode=FullScreen,
    }
\usepackage{color}
\usepackage{amsthm}
\usepackage{amsmath,amssymb}

\usepackage{xcolor}
\pagecolor{white}
\usepackage[english]{babel}
\usepackage[normalem]{ulem}

\usepackage{calrsfs}

\usepackage{animate}

\newcommand{\had}{\hat{a}^\dagger}

\newcommand{\hadmu}{\hat{a}^{\dagger \mu}}

\newcommand{\ha}{\hat{a}}

\newcommand{\ket}[1]{\left| #1 \right\rangle}
\newcommand{\bra}[1]{\left\langle #1 \right|}
\newcommand{\braket}[2]{\left\langle #1 | #2 \right\rangle}

\newcommand{\expect}[1]{\left\langle#1\right\rangle}
\newcommand{\eea}{\end{eqnarray}}
\newcommand{\bea}{\begin{eqnarray}}
\newcommand{\ee}{\end{equation}}
\newcommand{\be}{\begin{equation}}
\newcommand{\beq}{\begin{equation}}
\newcommand{\eeq}{\end{equation}}
\newcommand{\beqnn}{\begin{equation*}}
\newcommand{\eeqnn}{\end{equation*}}
\newcommand{\bes} {\begin{subequations}}
\newcommand{\ees} {\end{subequations}}

\newcommand{\sch}{Schr\"{o}dinger }

\AtBeginDocument{%
    \newwrite\bibnotes
    \def\bibnotesext{Notes.bib}
    \immediate\openout\bibnotes=\jobname\bibnotesext
    \immediate\write\bibnotes{@CONTROL{REVTEX41Control}}
    \immediate\write\bibnotes{@CONTROL{%
    apsrev41Control,author="08",editor="1",pages="1",title="0",year="1"}}
     \if@filesw
     \immediate\write\@auxout{\string\citation{apsrev41Control}}%
    \fi
}%

\begin{document}
\title{Decoherence Limits the Cost to Simulate an Anharmonic Oscillator}

\author{Tzula B. Propp}
\thanks{Current affiliation: QuTech, Delft University of Technology, Lorentzweg 1, 2628 CJ Delft, The Netherlands}
\affiliation{Center for Quantum Information \& Control, University of New Mexico, Albuquerque, NM 87131, USA}
\author{Sayonee Ray}
\thanks{Current affiliation: IonQ Inc, 4505 Campus Dr, College Park, MD 20740, USA}
\affiliation{Center for Quantum Information \& Control, University of New Mexico, Albuquerque, NM 87131, USA}
\affiliation{Department of Physics and Astronomy, University of Waterloo, Ontario, N2L 3G1, Canada}
\author{John B. DeBrota}
\affiliation{Center for Quantum Information \& Control, University of New Mexico, Albuquerque, NM 87131, USA}
\author{Tameem Albash}
\affiliation{Center for Quantum Information \& Control, University of New Mexico, Albuquerque, NM 87131, USA}
\affiliation{Department of Electrical and Computer Engineering, University of New Mexico, Albuquerque, NM 87131, USA}
\affiliation{Department of Physics and Astronomy, University of New Mexico, Albuquerque, NM 87131, USA}
\author{Ivan Deutsch}
\affiliation{Center for Quantum Information \& Control, University of New Mexico, Albuquerque, NM 87131, USA}

\affiliation{Department of Physics and Astronomy, University of New Mexico, Albuquerque, NM 87131, USA}

\begin{abstract}
We study how decoherence increases the efficiency with which we can simulate the quantum dynamics of an anharmonic oscillator, governed by the Kerr effect.  As decoherence washes out the fine-grained subPlanck structure associated with phase-space quantum interference in the closed quantum system, open quantum dynamics can be more efficiently simulated using a coarse-grained finite-difference numerical integration. We tie this to the way in which decoherence recovers the semiclassical truncated Wigner approximation (TWA), which strongly differs from the exact closed-system dynamics at times when quantum interference leads to cat states and more general superpositions of coherent states.  The regression in quadrature measurement statistics to semiclassical dynamics becomes more pronounced as the initial amplitude of the oscillator grows, with implications for the quantum advantage that might be accessible as system size grows in noisy quantum devices. Lastly, we show that this regression does not have the form of a convex noise model, such as for a depolarizing noise channel. Instead, closed quantum system effects interact with the open system effects, giving rise to distinct open system behavior.
\end{abstract}

\maketitle

\section{Introduction}

The macroscopic world is largely described by classical statistical physics even though the underlying fundamental description is quantum mechanical. As emphasized in the seminal work of Zurek \cite{Zurek1981,Zurek1991,Zurek1994,Anglin1997,Zurek1998,Paz1999,Karkuszewski2002,Zurek2001},
decoherence helps us understand the transition from the quantum-to-classical world: classical states are more robust to the coupling to the environment whereas highly nonclassical states, such as macroscopic superposition ``cat states" \cite{Dodo1974}, are fragile in the face of decoherence \cite{Leggett1987,Caldeira1983,Kim1992,Zurek2001}.

This observation not only explains why quantum effects are largely unobserved in the classical world, but it also represents the fundamental challenge of large scale quantum information processing as decoherence limits the quantum complexity one can harness at the macroscopic scale. While in principle one can tame decoherence through quantum error correction \cite{shor1995,Gottesman2001,eczoo}, we do not yet have the means to do so fault-tolerantly \cite{Sho1996,Aha1997,Kit1997,Kni1998} except at very small scales \cite{google2023}.  

Nevertheless, one may hope to achieve a meaningful quantum advantage without fault tolerant error correction.
A key question, thus, is how much complexity one can expect to harness in a noisy-intermediate scale quantum (NISQ) \cite{Preskill2018} device given that decoherence largely washes out the most nonclassical features.

One way to quantify the quantum-to-classical transition is using the phase space representation of quantum mechanics \cite{Curtright2013}.  The nonclassical features of quantum states are reflected in the negativity of the Wigner quasiprobability distribution \cite{Cahill1969,Hudson1974,Kenfack2004}. This is associated with the fine-grained subPlanck-scale structure that arises in quantum dynamics beyond what is described by classical flow in phase space \cite{Zurek2001}.  These properties challenge the classical simulation of macroscopic quantum dynamics; direct numerical propagation of the quasiprobability function scales poorly when trying to capture ultra-fine features \cite{Heller1976}, and applying techniques such as the methods of characteristics \cite{Weedbrook2012} or Monte Carlo sampling \cite{Sellier2015} is difficult when the quasiprobability function develops substantial negativity \cite{Welland2020}.  Decoherence, however, washes out subPlanck structure and the associated negativity by introducing a time scale over which the coherence of a quantum state is lost \cite{Habib1998}.  This transition occurs more rapidly as the system becomes more macroscopic \cite{Zurek2001}.  Thus, the phase space representation exhibits the trade off between robustness and complexity --- more subPlanck structure and negativity lead to faster decoherence, while those features that are robust to decoherence are essentially classical and more efficiently simulatable. 

In this work we consider how the phase space representation quantifies the robustness versus complexity trade off in the context of quantum simulation. Instead of focusing on negativity itself (which precludes semiclassical sampling), we will study how decoherence simplifies numerical simulation by erasing fine-grained structure. We consider a canonical toy model --- an anharmonic oscillator such as a single bosonic mode evolving under a Kerr nonlinearity. While a trivial model, it exhibits a wide range of well-studied nonclassical features, such as squeezing at short times \cite{milburn1986,Milburn1986b}, the collapse and revival of quantum oscillations \cite{polkovnikov}, and the generation of superpositions of coherent states~\cite{Miranowicz1990,tara1993,vanenk2003,vanenk2005, stobinska2008}.  These effects have been studied in a variety of platforms {including the seminal proposal in quantum optics~\cite{Yurke1986} and studies in trapped ions~\cite{Stobinska2011} as well as the realizations in atomic BEC~\cite{Greiner2002} and the pioneering experiments in circuit QED~\cite{Kirchmair2013} which now form the basis of quantum error-correcting codes that have been realized~\cite{Grim2020}.}  While there are  closed form solutions for this integrable model \cite{tara1993,vanenk2003, SUDHEESH2004}, including for the open quantum system~\cite{Milburn1986b, Peinova1990, Chaturvedi1991, Chaturvedi1991b, McDonald2022, McDonald2023}, in practice determining the expectation values of some observables can be numerically intensive when the mean number of bosons is deep in the macroscopic regime.  We emphasize that formally nothing is computationally complex in this model, as all quantities can be extracted with algorithms that scale polynomially with the mean boson number. Nonetheless, there remains a tradeoff between robustness and ease of numerical simulation. 

Our goal is to study how decoherence enables a more efficient simulation by quantifying the way in which coarse graining in phase space leads to efficient numerical integration of quantum dynamics. We do this by improving upon the methods of Ref.~\cite{stobinska2008} and simulating the system evolution at various coarse graining.  This helps us to understand the important role of representation in the efficiency of simulation.  While a representation of an open quantum system in terms of the Lindblad superoperator would indicate a less efficient representation of the open versus closed system, coarse graining of phase space allows us to retain the relevant information and reduce the complexity, as one would expect as the system becomes more ``classical." 

The remainder of this article is structured as follows. In Sec.~II we derive the dynamics of the quantum state under pure unitary evolution, both in discrete and continuous variable descriptions. In Sec.~III we expand on the continuous variable dynamics to include open system effects using the Fokker-Planck equations. In Sec. IV we study the time evolution of moments of the Wigner function which capture fine-grained information and quantify the cost of simulating the open quantum system when compared to closed system dynamics.  We use to this to better understand the nature of the  quantum-to-classical transition when compared to a trivial noise model where classical noise is added to a quantum signal, such as the depolarizing channel often used in the modeling of NISQ devices. We conclude with a summary and outlook in Sec.~V.

\section{Closed system dynamics}
We consider an anharmonic oscillator for a single bosonic mode with a Kerr nonlinearity, governed by the Hamiltonian
\begin{equation}\label{Hamiltonian}
\hat{H} =\frac{\kappa}{2} \hat{a}^{\dag 2} \hat{a}^{2} = \frac{\kappa}{2} \hat{n}(\hat{n}-1).
\end{equation}
Here and throughout $\hbar=1$.  This Hamiltonian arises, e.g., in quantum optics in the presence of an intensity-dependent index of refraction (Kerr effect)~\cite{Walls2008}, and in atom optics for bosonic atoms undergoing cold collisions \cite{Steel1998}.  For concreteness, we will consider the bosons here to be photons. The dynamics of this system has been well studied, and we review the salient results here. 
\begin{figure*}
	\includegraphics[width=1\linewidth]{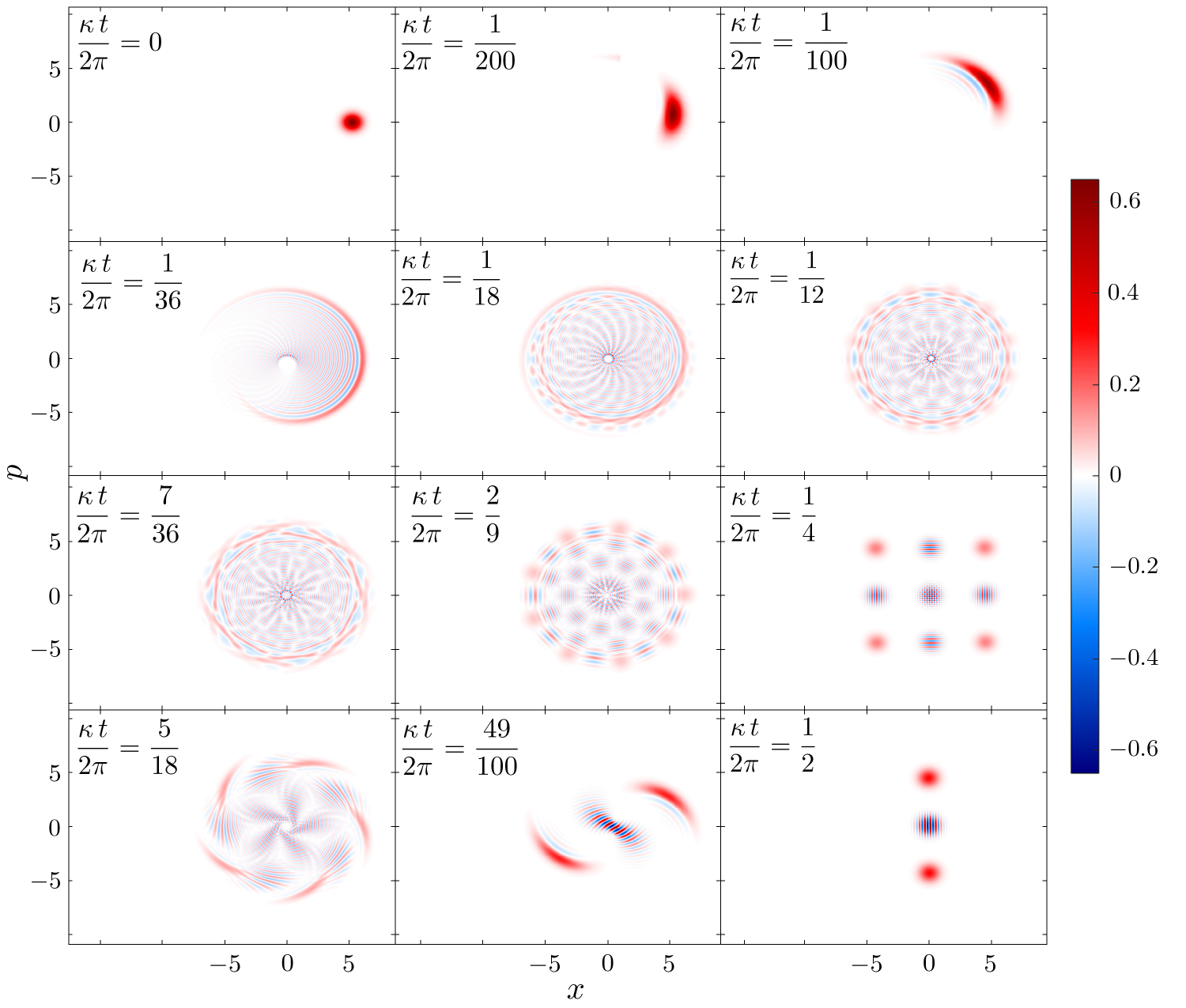} 
     \caption{The Wigner function for an initial coherent state with amplitude $\alpha_0=6$ at $t=0$ evolving under the Kerr interaction.  At short times we see characteristic squeezing by at longer times we see negativity in the Wigner function and the formation of intricate subPlanck-scale structure. At times $\kappa t_{M,N} =(M/N)2\pi$, $M$ and $N$ coprime, the state is a superposition of $N$ coherent states, Eq.~\eqref{KerrState}.  For $N \gg \alpha_0$, the coherent states have substantial overlap and we observe structures like the short time ``banana'' with minimal negativity in the top row. At later times, non-overlapping superpositions form. These include high-order kitten states which may appear multiple times (e.g. the $N=9$, $M=2$ kitten state in the center of the third row) as well as the \sch cat state at $t=\pi/\kappa$ (bottom right corner). Right before a kitten state forms, we observe shuriken-like Wigner functions (bottom left corner). In between $t=\pi/\kappa$ and $t=2\pi/\kappa$, the evolution is symmetric to the evolution between $t=0$ and $t=\pi/\kappa$ as the state progresses through the same states in reverse (not pictured for brevity). An animation depicting the continuous dynamics is given in~\cite{WignerVideo}.}\label{kittenstates}
\end{figure*}

The Heisenberg equation of motion for the annihilation operator $d\ha/dt =-i\kappa \had \ha \ha $
is integrable, $\ha (t) = e^{-i\kappa t \had \ha} \ha (0)$.  In the classical (mean field) limit, $\ha (t) \rightarrow \alpha_c(t)$, $\alpha_c(t) = e^{-i\kappa t |\alpha_c|^2}\alpha_c (0)$, describing the rotation of the phasor at an angular rate proportional to the amplitude squared (the classical Kerr effect). An arbitrary pure state evolves according to the trivial solution
\begin{equation}\label{psistate}
\ket{\psi (t)} =\sum_{n=0}^{\infty} c_n e^{-i\kappa t n(n-1)/2} \ket{n}.
\end{equation}
We will be particularly interested in the case that the initial state is a coherent state $\ket{\alpha_0}$, in which case $c_n = e^{-\frac{|\alpha_0|^2}{2}} \alpha_0^n/\sqrt{n!}$.\footnote{Similarly, one can also consider an initial squeezed state \cite{Banerjee1993}.} 

The evolution of the state is periodic, returning to its initial condition when $\kappa t = M\,2 \pi$, for integer $M$. There are additional revivals due to quantum interference at times $\kappa t_{M,N} =(M/N)2\pi$, where $M$ and $N$ are co-prime integers. This is derived in detail in Ref.~\cite{tanas2003}, and we only briefly summarize the method here. Upon substituting $t_{M,N}$ into Eq. (\ref{psistate}), one notes the $2N$-periodicity of $e^{-i\kappa t_{M,N} n(n-1)/2} $ so that $\ket{\psi (t_{M,N})}$ has the form of a generalized coherent state \cite{bialynickabirula1968}. Performing a discrete Fourier transform and, after minor rearrangement, we arrive at the form,
\bea\label{KerrState}
\ket{\psi(t_{M,N})} &=&  \sum_{k=0}^{2N-1}f_k \ket{\alpha_0  e^{ik\frac{\pi}{N}}} ,\nonumber \\
\mathrm{where}\; \; f_k &=& \frac{1}{2N} \sum_{n=0}^{2N-1} e^{-i\frac{\pi}{N}[nk-Mn(n-1)]}
\eea 
One can show that $N$ values of $f_k$ will be nonzero for any particular choice of (co-prime) $N$ and $M$, and  $|f_k|^2=1/N$~\cite{tanas2003}. Therefore, at the discrete times $t_{M,N}$, the state revives to coherent superposition of $N$ copies of the initial state, distributed in phase space on the circle with radius $|\alpha_0|$ at multiples of $2\pi/N$.

For the case $M=1$, $N=2$, one obtains $\ket{\psi(t_{1,2})} = \frac{1}{\sqrt{2}}(\ket{i \alpha_0}+i\ket{-i \alpha_0})$, a ``cat state." For general $N$ and $M$, the state is a superposition of $N$ coherent states which we will denote ``kitten states" as defined in \cite{tanas2003}.\footnote{The kitten states and the cat state generated by the Kerr interaction all have a remarkable feature: each phase is such that the overall normalization constant is independent of the coherent state amplitude $\alpha_0$, which is not the case for a general superposition of coherent states due to their non-orthogonality.} These states form a dense set and exist at all $t_{M,N}$, but only come to the fore when the coherent states are well-separated. 

We are particularly interested in studying the dynamics of the system using phase space representations.  The equation of  motion for the Wigner function, $W(\alpha,\alpha^*,t)$, with a quadratic nonlinearity is
\begin{eqnarray}\label{FPClosed}
  \frac{\partial W}{\partial t} &=& \{H_W,W\}_{MB} \nonumber \\ 
  &=&  -i \left( \frac{\partial H_W}{\partial \alpha} \frac{\partial W}{\partial \alpha^*} - \frac{\partial H_W}{\partial \alpha^*} \frac{\partial W}{\partial \alpha}\right)  \\
   && -\frac{i}{8} \left( \frac{\partial^3 H_W}{\partial^2 \alpha^* \partial \alpha} \frac{\partial^3 W}{\partial^2 \alpha \partial \alpha^*}  -\frac{\partial^3 H_W}{\partial^2 \alpha \partial \alpha^*}\frac{\partial^3 W}{\partial^2 \alpha^* \partial \alpha}\right) \nonumber,
\end{eqnarray}
where $H_W(\alpha,\alpha^*) = \frac{\kappa}{2}|\alpha|^2(|\alpha|^2-2)+\frac{\kappa}{4}$ is the Weyl-symbol of the Hamiltonian and $\{H_W,W\}_{MB}$ is the Moyal bracket~\cite{Polkovnikov2013}.  The first line in Eq.~\eqref{FPClosed} is  $\{H_W,W\}_{PB}$, the Poisson bracket, corresponding to the classical flow on phase space generated by $H_W$.  The second line represents the nonclassical dynamics, which can lead to negativity of the Wigner function.  These terms become nonnegligible when the Wigner function develops subPlanck-scale structure as this corresponds to a rapidly varying function whose higher order derivatives are at least order one.  Neglecting these quantum dynamics is known as the truncated Wigner approximation (TWA) \cite{Polkovnikov2013}, which well describes the dynamics for times short times compared to the Ehrenfest time (that is, when corrections to the Poisson bracket become nonnegligible).

Substituting $H_W$ for the Kerr Hamiltonian into Eq.~\eqref{FPClosed}, we have
\begin{eqnarray}
  \frac{\partial W}{\partial t}
  &=&  -i\kappa(|\alpha|^2-1) \left( \alpha^* \frac{\partial W}{\partial \alpha^*} - \alpha \frac{\partial W}{\partial \alpha}\right)\nonumber\\
   && -\frac{i\kappa}{4} \left( \alpha \frac{\partial^3 W}{\partial^2 \alpha \partial \alpha^*}  - \alpha^*\frac{\partial^3 W}{\partial^2 \alpha^* \partial \alpha}\right) .
\end{eqnarray}
The first line is the TWA representing classical flow on phase space.  Note $ -i(\alpha^* \partial_{ \alpha^*}  - \alpha \partial_{ \alpha})  = X \partial_P -P \partial_X $ is the rotation operator on phase space, where the quadratures are defined by $\alpha = (X+iP)/\sqrt{2}$.  The TWA has the expected form of a Kerr effect, that is, a rotation at angular rate $\kappa|\alpha|^2$ for $|\alpha|$ large. The evolution according to the TWA can be solved by the method of characteristics, $W_{\mathrm{TWA}}(\alpha,\alpha^*,t)=W_{\mathrm{TWA}}(\alpha(-t),\alpha^*(-t),0)$, where $\alpha(t)=e^{-i\kappa t(|\alpha|^2-1)}\alpha$ is the classical flow.  The TWA evolution of an initial coherent state represented as a Gaussian wavepacket is thus, $W_{\mathrm{TWA}}(\alpha,\alpha^*,t) = \frac{1}{\pi}\exp\{-2|e^{i\kappa t(|\alpha|^2-1)}\alpha-\alpha_0|^2\}$, as shown in the top row of Fig.~\ref{TWAState}. For short times the nonlinear rotation ``shears" the distribution, leading to a squeezed Gaussian.  At longer times the distribution becomes stretched and becomes highly nonGaussian, but remains a positive probability distribution in the TWA.  At these longer times the corrections to the TWA become nonneglegible.  Figure~\ref{kittenstates} shows the Wigner function evolution for $\alpha_0=6$, calculated using the numerical methods described in Appendix \ref{app:NumericalMethods}.  For very short times, the TWA evolution approximately matches the exact evolution, but negativity soon develops and revivals occur at the expected times, yielding kitten states and subPlanck structure.  

While in principle the exact state is available in any representation, in practice extracting measured values of observables becomes numerically intensive when $\alpha_0$ is sufficiently large.
We will specifically focus on symmetrically-ordered correlation functions, which are moments of the Wigner function,
\begin{equation}\label{marginalXP} 
\langle \{\hat{X}^n\hat{P}^m\}_{\mathrm{sym}}\rangle(t) = \int dX dP \; X^n P^m\; W(X,P,t),
\end{equation}
where the symmetrically-ordered product of $\hat{X}$ and $\hat{P}$ is to be normalized by $n+m$, e.g., $\{\hat{X}\hat{P}\}_{\mathrm{sym}} = \frac{1}{2} (\hat{X}\hat{P}+\hat{P}\hat{X})$. 
Such correlation functions arise, e.g., in considering moments of a quadrature $\hat{X}_\theta= \cos\theta \hat{X} +\sin\theta \hat{P}$ optically measured in homodyne detection, $\langle \hat{X}^n_\theta \rangle (t) = \int dX_\theta \mathcal{P}_\theta(X_\theta,t) X^n_\theta$. 

For the closed system, the probability distribution which determines these moments $\langle \hat{X}^n_\theta \rangle (t)$ is the square of the wave function as a function of the quadrature $X_\theta$ eigenvalue, which has an analytic form
 \begin{eqnarray}\label{marginaldefearly}
&    \mathcal{P}(X_\theta, t)= |\langle X_\theta |\psi(t)\rangle |^2 &\\
 &= \left | \sum_{n=0}^\infty e^{-\frac{|\alpha_0|^2}{2}} \frac{\alpha_0^n}{\sqrt{n!}} e^{-i\left(\frac{\kappa t}{2} + \theta\right)} e^{-i\frac{\kappa t n^2}{2}} u_n(X_\theta)\right|^2, &\nonumber  
 \end{eqnarray}
where we have defined $u_n(X_\theta)  = A_n H_n(X_\theta) e^{-\frac{X_\theta^2}{2}}$, with $A_n = \left(\sqrt{\pi} 2^n n!\right)^{-1/2}$ and $H_n(X_\theta)$ the $n$th Hermite polynomial. The relevant number of terms to evaluate this formal expression grows as $\sim |\alpha_0|^4$. More generally, this probability distribution can be obtained by marginalizing the Wigner function
\begin{equation}
\mathcal{P}(X_\theta, t) = \int dP_\theta W(X_\theta, P_\theta,t)
 \end{equation}
where we define $P_\theta = X_{\theta+\pi/2}$.
 
The higher order moments of the marginal capture the fine-grained structure in the Wigner function, and generically this is hardest to simulate in the large $|\alpha_0|$ limit, when nonclassical subPlanck-scale structure develops.  However, as discussed above, this fine grained structure is washed out by decoherence. Thus, our goal is to determine how much of this complexity remains in the open quantum system and how the reduction of this complexity leads to more efficient simulations.

The Kerr anharmonic oscillator is a useful system for benchmarking the nonclassical dynamics and comparing numerical simulations to the exact analytic solution.  
The conservation of photon number implies normally ordered correlation functions have a simple closed form for an initial coherent state,
\begin{equation}
\langle \hat{a}^{\dagger\mu} \ha^\nu \rangle (t) \equiv \langle \alpha_0 | \hat{a}^{\dag \mu}(t) \hat{a}^\nu (t) | \alpha_0 \rangle,
\end{equation}
where $\mu$ and $\nu$ are nonnegative integers.
Using $\hat{a}^\nu(t)\ket{\alpha_0} = \alpha_0^\nu e^{-i\frac{\kappa t}{2}\nu(\nu-1)} \ket{\alpha_0 e^{-i\nu\kappa t}}$  and $\langle \alpha | \beta \rangle = e^{-\frac{1}{2}|\alpha - \beta |^2} e^{-\frac{1}{2}(\alpha\beta^* - \alpha^*\beta)}$  one finds
\begin{eqnarray}\label{normallyorderedexp}
\langle \hat{a}^{\dagger\mu} \ha^\nu \rangle (t) = (\alpha_0^*)^\mu(\alpha_0)^\nu e^{i\frac{\kappa t}{2}[\mu(\mu-1)-\nu(\nu-1)]} \nonumber \\ 
\times \;e^{-|\alpha_0|^2(1-\cos[(\mu-\nu)\kappa t])}e^{i|\alpha_0|^2\sin[(\mu-\nu)\kappa t]},
\end{eqnarray} 
which tends towards zero exponentially with $|\alpha_0|^2$ except at special times when $\kappa t = M (\mu-\nu)\pi$, with $M$ an integer. Understanding quantum dynamics at these ``recurrences'' will be the focus of Sec.~\ref{sec:QuadMoments}.


Symmetrically ordered expectation values are more costly to simulate, as they require calculating multiple normally-ordered expectation values. They are related to the normally ordered ones according to
\begin{eqnarray}
\label{expectschars}
\expect{\{\hadmu \ha^\nu\}_{\mathrm{sym}}}&= &\\
&& \hspace{-2cm} \sum_{\gamma=0}^{{\rm min}(\mu,\nu)}\frac{\mu!\nu!}{\gamma!(\mu-\gamma)!(\nu-\gamma)! 2^\gamma} \expect{\hat{a}^{\dagger \mu-\gamma} \,\hat{a}^{\nu-\gamma}}. \nonumber 
\end{eqnarray} 
See Appendix \ref{app:symtonorm} for a derivation of this expression. For very large order moments with large $|\alpha_0|$, accurate calculation of such sums becomes numerically intensive.

\section{Open System dynamics}
We consider the simplest model of decoherence for the single mode --- the damped simple harmonic oscillator due to photon loss, whereby the state evolves according to the master equation
\begin{equation}\label{mastereq}
\frac{\partial \hat{\rho}}{\partial t} = -i[\hat{H},\hat{\rho}]+\mathcal{L}[\hat{\rho}],
\end{equation}
where 
\begin{equation}\label{ampdampLindbladian}
\mathcal{L}[\hat{\rho}] = -\frac{\gamma}{2}\left(\had\ha\hat{\rho}+ \hat{\rho}\had\ha\right) + \gamma \ha \rho \had
\end{equation}
is the Lindbladian describing photon loss in a zero temperature reservoir. The extension to a finite temperature reservoir is straightforward but does not qualitatively change any of the conclusions we draw below.\footnote{This is because we study the difference between exact open quantum system dynamics and dynamics given by the TWA, and the latter also can include finite-temperature effects.} 
In the absence of Hamiltonian evolution, for a general superposition of coherent states, $\ket{\psi}=\sum_k f_k \ket{\alpha_k}$ we have the closed form solution~\cite{Walls2008}
\begin{eqnarray}\label{decaykittengen}
e^{\mathcal{L}t}[\ket{\psi}\bra{\psi}] &=& \sum_k |f_k|^2 \ket{\alpha_k e^{-\frac{\gamma t}{2}}}\bra{\alpha_k e^{-\frac{\gamma t}{2}}} \nonumber\\
&& \hspace{-2cm} + \sum_{k\neq k'} f_k f^*_{k'} \braket{\alpha_{k'}}{\alpha_k}^{1-e^{-\gamma t}} \ket{\alpha_{k'} e^{-\frac{\gamma t}{2}}}\bra{\alpha_k e^{-\frac{\gamma t}{2}}}.\nonumber\\
\end{eqnarray}
The amplitude of coherent states in the mixture decay as expected, but the coherences in the superposition of coherent states decay more rapidly, at a rate that depends on their overlap. In particular, for a cat state, for $t\ll 1/\gamma$,
\begin{equation}
\braket{-\alpha_0}{\alpha_0}^{1-e^{-\gamma t}} \approx \braket{-\alpha_0}{\alpha_0}^{\gamma t}= e^{-2|\alpha_0|^2 \gamma t},
\end{equation}
and this coherence decays at the rate $2|\alpha_0|^2\gamma$. For macroscopic $\alpha_0$ this implies that the coherence between the coherent states is lost essentially instantaneously compared to the time of substantial energy loss.

The Wigner function for the cat state thus evolves under the damping channel as
\begin{eqnarray}\label{CatWigner}
W(X,P,t) &=& \frac{1}{2}W_0(X,P-X_0 e^{-\frac{\gamma t}{2}}) \nonumber\\
&& \hspace{-2.25cm}+ \frac{1}{2}W_0(X,P+X_0 e^{-\frac{\gamma t}{2}})+ e^{-X_0^2 \gamma t}\sin(2X X_0)W_0(X,P), \nonumber \\
\end{eqnarray}
where $W_0(X,P) = e^{-(X^2+P^2)}/\sqrt{\pi}$ is the Wigner function of the vacuum. The subPlanck-scale structure decays rapidly, at the rate $X_0^2\gamma = 2\expect{\hat{n}} \gamma$. This can also be seen directly from the phase space dynamics. Recall the Weyl representation of the Lindbladian
\begin{eqnarray}
\mathcal{L}[\hat{\rho}]_W &=& \frac{\gamma}{2} \left(\frac{\partial}{\partial \alpha}\alpha  + \frac{\partial}{\partial \alpha^*}\alpha^* \right)W+  \frac{\gamma}{2} \frac{\partial^2 W}{\partial \alpha \partial \alpha^*} \\
 && \hspace{-0.75cm} =  \frac{\gamma}{2} \left(\frac{\partial}{\partial X}X  + \frac{\partial}{\partial P}P \right)W+  \frac{\gamma}{4} \left( \frac{\partial^2W}{\partial X^2} + \frac{\partial^2W}{\partial P^2} \right) , \nonumber 
\end{eqnarray}
which is a Fokker-Planck equation. The first order derivative ``drift" terms generate the decay of energy of the coherent state. The Laplacian diffusion terms lead to the rapid washing out of subPlanck scale structure, and thus decoherence.  Indeed, we see the action of diffusion on the interference term in Eq.~\eqref{CatWigner} gives, to leading order in $X_0$, $\frac{\gamma}{4}(\frac{\partial^2}{\partial X^2 }+\frac{\partial^2}{\partial P^2})\sin(2XX_0)W_0(X,P) \approx -X_0^2\gamma\sin(2XX_0)W_0(X,P)$, which shows that diffusion leads to decays of coherence with the same rate we obtained by different methods. The finer the subPlanck scale structure, the faster decoherence washes it out~\cite{Zurek2001} -- a manifestation of the robustness vs.\ complexity tradeoff.

\begin{figure*}[t] 
	\includegraphics[width=\linewidth]{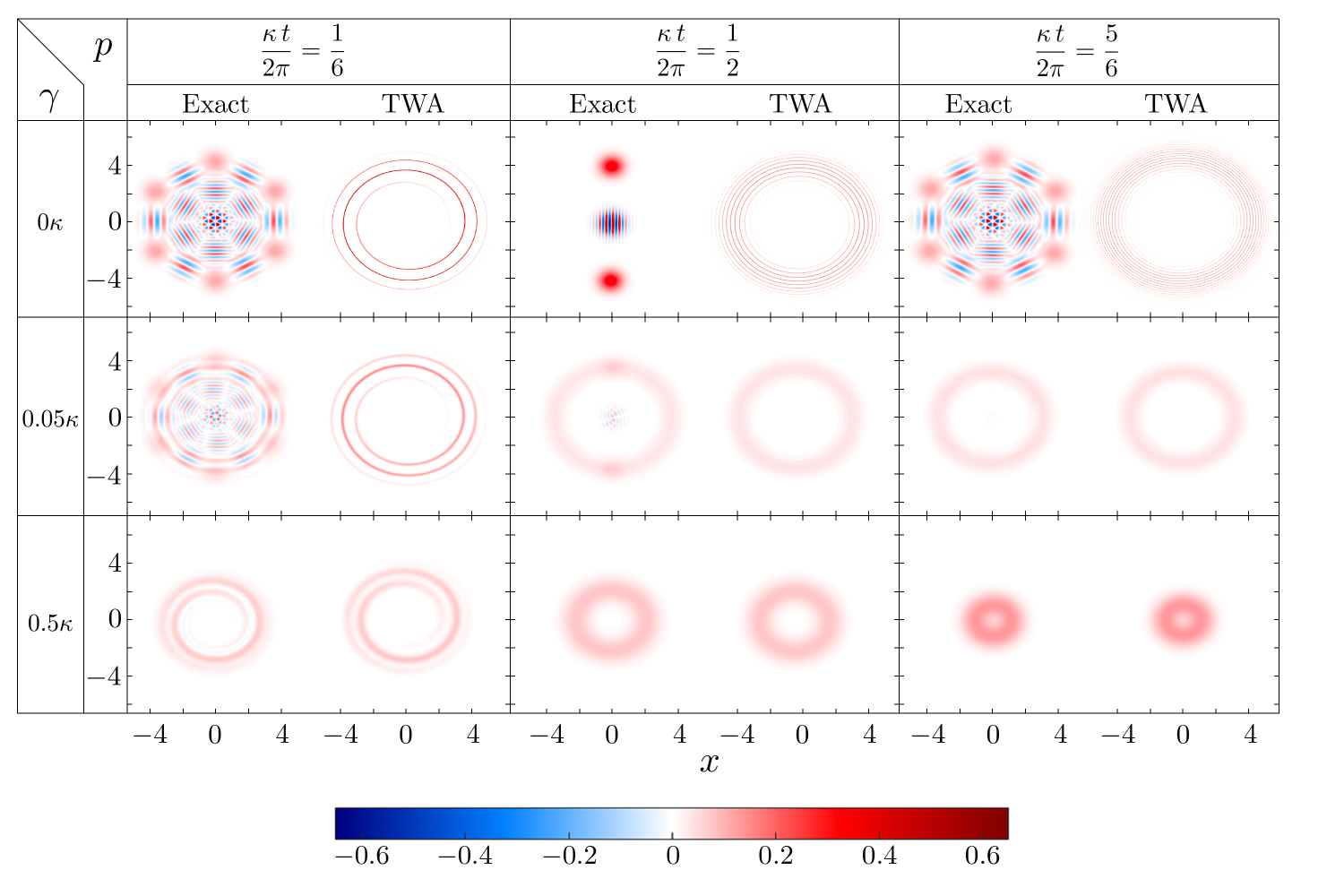} 
	\caption[]{The Wigner functions of the closed system (top row) and open system (bottom two rows), evolving from an initial coherent state with $\alpha_0=4$, are plotted at times corresponding to the first six-kitten state (first column), the cat state (middle column), and the second six-kitten state (last column). In each inset, both the exact quantum state and the state calculated using the truncated Wigner approximation (TWA) are shown on the left and right, respectively. For nonzero coupling to the environment, the exact state more closely resembles that given by the TWA, with the resemblance increasing with both decoherence strength $\gamma$ (vertical axis) and time (horizontal axis).}
	\label{TWAState}
\end{figure*}

The equation of motion for the Wigner function under the concurrent action of the Kerr Hamiltonian and photon loss is given by $\frac{\partial W}{\partial t} =\{H,W\}_{MB}+\mathcal{L}[\hat{\rho}]_W$.  While a formal solution exists for the Husimi representation~\cite{Milburn1986b}, no such solution exists for the Wigner function, and generally one must resort to numerical integration.  As first studied by Stobinska {\em et al.}~\cite{stobinska2008}, given the invariance of the Kerr interaction under rotation in phase space, this is best done in polar coordinates by expressing the complex amplitude $\alpha \equiv r e^{i \phi}$, giving the equation of motion 
\begin{widetext}\bea\label{FokkerPlanck}
\frac{\partial}{\partial t} W(r,\phi, t)&=&\! \left[\kappa (r^2 - 1)\frac{\partial}{\partial\phi} - \frac{\kappa}{16}\left(\frac{1}{r} \frac{\partial}{\partial r} + \frac{\partial^2}{\partial r^2} + \frac{1}{r^2} \frac{\partial^2}{\partial\phi^2} \right)\frac{\partial}{\partial\phi} \right] W(r,\phi,t)\nonumber \\
&+& \left[\gamma\left(1 + \frac{ r}{2} \frac{\partial}{\partial r}\right) + \frac{\gamma}{8} \left(\frac{1}{r} \frac{\partial}{\partial r}+\frac{\partial^2}{\partial r^2 } + \frac{1}{r^2}\frac{\partial^2}{\partial\phi^2}\right)\right]W(r,\phi,t).
\eea\end{widetext}
The first line is the closed-system evolution, where the first term is the expected classical rotation in phase phase depending on the amplitude squared, $r^2$ (the TWA evolution), and the second term is the nonclassical Hamiltonian flow.  The second line is the Fokker-Planck equation where the first term is the damping (drift radially inward to the origin), and the second term is diffusion. 

Closed form solutions for the Wigner function are generally not available.  Recently McDonald and Clerk developed a powerful method based on diagonalization of the Lindblad superoperator, which follows from fundamental dissipative symmetry present in all quadratic bosonic Lindbladians~\cite{McDonald2023}. Using this they were able to obtain a closed form propagator in phase space, and a solution for the Wigner function in terms of a compact but infinite sum.  We seek to probe the role of dissipation in reducing computational complexity associated with subPlanck-scale structure.  For this we look to the the approach of  
 Stobinska {\em et al.}, who used a finite difference method in order to numerically integrate this partial differential equation (PDE)~\cite{stobinska2008}.  This method discretized the Wigner funcion in phase space but the numerics were limited to small $\alpha_0$ as the required grid size would otherwise grow too large.  This is a reflection of the need to capture the fine grained structure that develops in the Wigner function.  However, as noted, we expect the fine-grained structure to be limited by diffusion (decoherence), and thus a coarse grained approximation to the Wigner function should give a good approximation for the open quantum system. Similarly, since decoherence is washing out the features generated by nonclassical flow, we expect the TWA to give a better representation of the state in the presence of decoherence as seen in Fig.~\ref{TWAState}. We study both of these quantitatively in the section to follow. 

\section{Quadrature Moments} \label{sec:QuadMoments}
We study in detail the behavior of expectation values of powers of the phase quadrature operator 
\begin{equation}
 \langle \hat{X}^n_\theta \rangle (t) = \sum_{m=0}^{n}{n \choose m} \cos^{n-m}\theta \sin^m\theta\langle \{\hat{X}^{n-m} \hat{P}^m\}_{\mathrm{sym}} \rangle(t).   
\end{equation}
 These describe statistics of homodyne measurements, with higher-order moments corresponding to finer-scale features of the Wigner function, and with their evolution governed by the quantum dynamics. As discussed above, the distinction between the quantum and classical dynamics is due to quantum interference between different photons numbers. For an initial distribution of field amplitudes, as for an initial coherent state, under classical dynamics the quadrature moments will collapse as a function of time due to the nonlinear phase shifts, whereas under closed-system unitary dynamics these moments also exhibit a series of revivals at recurrence times due to quantum interference, as discussed above. Away from these times, the expectation value is well-described by the semiclassical dynamics given by the TWA~\cite{Polkovnikov2013}. 
 
 We seek to understand the deviation of the exact quantum dynamics from the semiclassical dynamics given by the TWA. In this section, we first derive the existence and times $t^{(n)}_{M,N}$ of these recurrences for the $n^{th}$ phase-quadrature moment. Then, we show why the deviations from the TWA occur \emph{solely} at times near these recurrences (with the effect becoming more prominent in the large-$\alpha_0$ limit), despite the dense set of highly-quantum kitten states throughout the evolution. Finally, we study the quantum behavior of $\langle \hat{X}^n_\theta \rangle (t)$ itself in terms of its deviation from semiclassical dynamics.

 The existence of recurrence times can be seen from the expression for normally ordered correlation functions for the closed system given an initial coherent state with amplitude $\alpha_0$ in Eq.~\eqref{normallyorderedexp}. We observe that the magnitude of $\langle \hat{a}^{\dag \mu} \hat{a}^\nu \rangle$ has periodicity (that is, recurrence) at times
\bea\label{period}
T_{\mu\nu}=\frac{2\pi}{|\mu-\nu|\kappa}.
\eea 
$T_{\mu\mu}$ is undefined, which is consistent with the Kerr interaction being number preserving; there is never a recurrence for powers of the number operator because it does not change during the evolution. Since the expectation value of any normally ordered operator can be expressed as a sum of expectation values of the form of Eq.~\eqref{normallyorderedexp}, any normally ordered operator will also exhibit recurrences at these times. Furthermore, the period $T_{\mu \nu}$ of an expectation value $\expect{\hadmu \ha^\nu} (t)$ is unchanged by operator ordering since the recurrence depends only on the difference $\mu-\nu$ and this difference is preserved by the commutation relations.  Because of this, a symmetrized expectation value $\expect{\{\hadmu \ha^\nu\}_{\rm sym}} (t)$ also has the same period $T_{\mu \nu}$.

The dependence of the recurrence times on the difference $\mu-\nu$ and not $\mu$ and $\nu$ separately is also true for the open quantum system governed by Eq. (\ref{mastereq}). The open system recurrences also exhibit independence of operator ordering, so that normally ordered and symmetrically ordered operators with the same difference $\mu-\nu$ have the same recurrence times. We derive these results below with further details given in Appendix \ref{app:reccurencethms}.

The Lindblad master equation Eq.~\eqref{mastereq} expressed in the number basis is given by
\begin{widetext}
 \bea\label{densityelementkerr}
\frac{d}{dt} \rho_{nm} (t) = - \left(\frac{\gamma}{2}(n+m) + i\frac{\kappa}{2}\left( (n^2 -n) - (m^2-m)\right) \right) \rho_{nm} (t) + \gamma \sqrt{(n+1)(m+1)}  \rho_{n+1,\,m+1}(t). 
 \eea\end{widetext}
We can write a formal solution for the density matrix elements as a generalized discrete Fourier series
\bea\label{densityDiscreteFourier}
 \rho_{nm}(t) = \sum_{n',m'} A_{n'm'}^{nm} e^{-\tilde{\gamma}_{n'm'} t} \delta_{n'-m', n-m},
\eea 
where we define a complex  function
\bea
\label{gammajkCoeffs2}
\tilde{\gamma}_{n'm'} = \frac{\gamma}{2}(n'+m') + i\frac{\kappa}{2}\left( (n'^2 -n') - (m'^2-m')\right).
\eea
For the closed  system, $\tilde{\gamma}_{n'm'}$ is purely imaginary, and the Fourier coefficients $A_{n'm'}^{nm}$ are all zero except for $n'=m,m'=m$ with $A_{nm}^{nm}=\rho_{nm}(0)$. The sum thus collapses to a single term. For the open system, all Fourier coefficients with $n'<n$ and $m'<m$ are zero, and the coefficients $A_{n'm'}^{nm}$ can be solved recursively by truncating the Fock space (for details see Appendix \ref{app:reccurencethms}). 

In all cases the form of Eq.~\eqref{densityDiscreteFourier} implies that {each set of elements $\rho_{nm}(t)$ of the density matrix on a diagonal ``stripe," with $n-m$} a fixed constant, evolves independently. {This follows directly for the ``weak symmetry'' associated with the Lindbladian on the dissipative Kerr oscillator, recently identified by McDonald and Clerk~\cite{McDonald2022}.}  We can use this fact to study the nature of the recurrences in the open quantum system.  It follows from Eq.~(\ref{densityDiscreteFourier}) that  
\begin{align}
\label{expectsDirect}
\expect{\hadmu \ha^\nu} (t) &=\sum\limits_{nm} P_{nm}^{\mu\nu} \rho_{nm}(t) \delta_{n-m,\mu-\nu} \nonumber \\
&= \sum_{nm}
R_{nm}^{\mu\nu} e^{-\tilde{\gamma}_{nm} t} \delta_{n-m,\mu-\nu} , 
\end{align} 
where $P_{nm}^{\mu\nu}$ is a positive real combinatorial factor and where we have defined new generalized Fourier coefficients  
\bea\label{FourierCoeffs}
R_{nm}^{\mu\nu} = \sum_{n'} P_{n-m+n',n'}^{\mu\nu} A_{nm}^{n-m+n',n'}.
\eea 
From this form of the expectation value, we make the following three conclusions: 
\begin{enumerate}
\item The recurrence times of operators depend only on the difference, $\mu-\nu$, and not the powers $\mu$ and $\nu$ directly. 
\item The recurrence times are unchanged by operator ordering. For symmetricly ordered and normally ordered expectation values the combinatorial factor $P_{nm}^{\mu\nu}$ will be different, but otherwise the expressions for the expectation value in Eq.~(\ref{expectsDirect}) are the same. 
\item Markovian open system effects will not change the recurrence times as the correlation function depends on the same frequency components that contribute in the closed quantum system. 
\end{enumerate}

Thus, the periodicity of the expectation value is independent of operator ordering, and of being for an open or closed system, up to rescaling of those coefficients by the real part of $\tilde{\gamma}_{nm}$ and a change of the value of the combinatorial prefactor; both of these leave the recurrence time unchanged.\footnote{Note that this argument fails for $\nu=\mu$ since these depend on the diagonal of the density matrix, where $\tilde{\gamma}_{jk}$ is completely real. However, this is to be expected, as these expectation values are simply powers of the number operator and exhibit no periodicity/recurrence in the first place.} This behavior extends to any anharmonic, single-mode bosonic system at finite temperature as we show in Appendix~\ref{app:reccurencethms}.

\begin{figure*}[t] 
	\includegraphics[width=\linewidth]{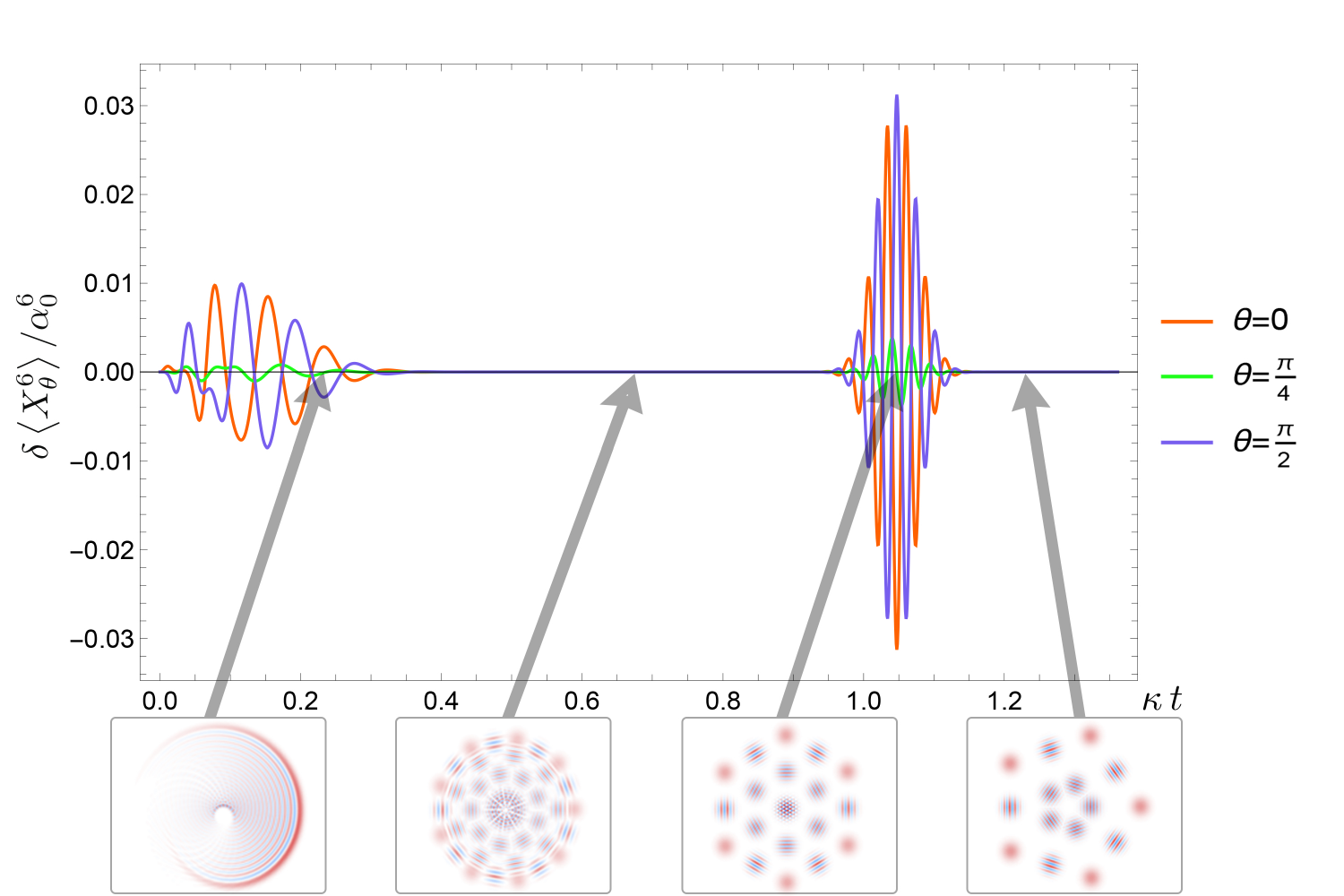} 
	\caption[]{Deviation of the expectation value $\langle\hat{X}_\theta^6\rangle(t)$ from its value calculated under the TWA is plotted for the closed quantum system evolving under the Kerr interaction (that is, $\delta \langle \hat{X}_\theta^6 \rangle (t)\equiv \langle \hat{X}_\theta^6 \rangle (t)- \langle \hat{X}_\theta^6 \rangle_{\rm TWA}(t)$). The initial state is a coherent state with $\alpha_0=6$, and we normalize the expectation value by $\alpha_0^6$. Different values of $\theta$ correspond to different phase quadratures and give the same times of recurrence. The $\theta$ are chosen to correspond to angles present in the $6$ kitten state, which form at $\kappa t=\pi/3$. Plots of the Wigner function at times $\kappa t = \pi/18,\, 2\pi/9, \pi/3$, and $2\pi/5$ are included as insets. Notably, there is a dense set of times throughout evolution where the Wigner function is a kitten state with highly quantum features, yet the TWA only fails to give the correct expectation value for $\hat{X}_\theta^6$ near the beginning of evolution and near the first recurrence time $t^{(6)}_{1,6}$. At the times $\kappa t =  2\pi/9$ and $\kappa t = 2\pi/5$, we do not observe a deviation from the TWA due to a mismatch between the symmetry of the measurement operator $\hat{X}_\theta^6$ (which has the $6$-fold symmetry of the cyclic group $\mathbb{Z}_6$) and the symmetries of the $5$ and $9$ kitten states ($\mathbb{Z}_5$ and $\mathbb{Z}_9$, respectively). Near the recurrence time at $\kappa t=\pi/3$, there are additional contributions from high order kittens such that the terms of order $\mathcal{O}(\braket{\alpha_0}{\alpha_0\,e^{i\frac{2\pi}{N}}}) $ in the last line of Eq.~\eqref{kittenexpectexpanded} become non-negligible, which are exponentially suppressed in the large-$\alpha_0$ limit.}
	\label{PhaseCurves}
\end{figure*}

To understand why away from these recurrences the system is well-described by the TWA, consider the normally ordered expectation value $\bra{\psi (t_{M,N})} \hadmu \ha^\nu \ket{\psi (t_{M,N})}$, where $\ket{\psi (t_{M,N})}$ is the $N$ kitten state at time $t_{M,N}$ defined in Eq.~\eqref{KerrState}. This expectation value is only sensitive to the number of kittens $N$ when $\mu-\nu = p\,N$ with $p$ a non-zero integer. To see this, note from its form that $\ha^N \ket{\psi (t_{M,N})} = \pm\alpha_0^N \ket{\psi (t_{M,N})}$, with the minus sign occurring for $N$-even and $M$-odd only.\footnote{This is due to an overall rotation in the location of kitten states when $N$ is even and $M$ is odd.} This property follows from the following considerations. Up to normalization and an overall phase, the effect of $\ha$ on an $N$-kitten state is to permute the phases of the coherent states in the superposition, forming a cyclic group ($\mathbb{Z}_N$) \cite{Nash2016}, so that all phases return to their original values after $N$ applications \cite{6kittenVideo}. Because of this property, these states were originally denoted ``generalized coherent states."~\cite{bialynickabirula1968}.
The same is true for the action of $\had$ on $\bra{\psi (t_{M,N})}$. Only when $\ha$ and $\had$ are applied a number of times whose difference is an integer multiple of $N$ do the phases fail to average out to zero, giving rise to recurrence at the times predicted by the period formula in Eq.~\eqref{period}. The recurrence becomes more sharply peaked at this time only in the large-$\alpha_0$ limit, when the coherent states are approximately orthogonal. 

\begin{figure*}[t]
\includegraphics[width=1\linewidth]{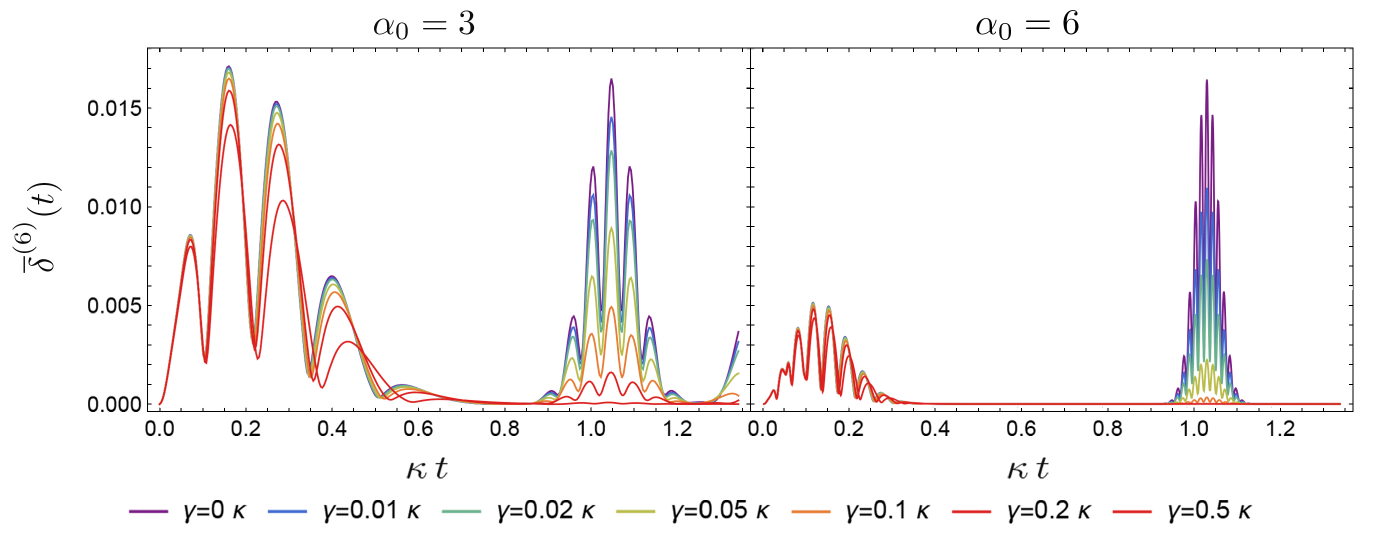} 
     \caption{Averaged-deviation from the TWA ($\overline{\delta}^{(n)}(t)$) for $\alpha_0=3$ (left) and $\alpha_0=6$ (right) as defined in Eq.~\eqref{InstantDeviationAverageDef}. In both cases, the effect of increased decoherence is to suppress quantum deviations from the TWA. Increasing the amplitude of the initial coherent state both magnifies the effects of decoherence and reduces the initial short-time deviation from the TWA, and the oscillations become more sharply peaked around the recurrence times $t^{(n)}_{M,N}$ \cite{polkovnikov,Polkovnikov2003}.
     }\label{AverageOscillationTime}
\end{figure*}

We derive these properties for general kitten states (not only those generated by the Kerr interaction) in Appendix \ref{app:expectkittens}. The behavior of the expectation value is given by 
\begin{widetext}\bea\label{kittenexpectexpanded}
\bra{\psi (t_{M,N})} \hadmu \ha^\nu \ket{\psi (t_{M,N})} = \begin{cases}
\big\langle\hat{n}\big\rangle^{\mu} &\,\mu=\nu,\\
\alpha_0^{*\mu-\nu} \big\langle\hat{n}\big\rangle^{\nu}&\,\mu-\nu=p\,N,\,p\in \mathbb{Z}^{\pm},\\
0+\mathcal{O}\bigr(\big\langle\alpha_0\big|\alpha_0e^{i\frac{2\pi}{N}}\big\rangle\bigr) & \mu-\nu=p\,N,\,p\notin \mathbb{Z},
\end{cases} \nonumber \\
\eea\end{widetext} 
where $\mathbb{Z}^{\pm}$ is the set of non-zero integers, and $\mathbb{Z}$ is the set of all integers including zero.  Note that, for normally ordered expectation values of a \emph{mixture} of $N$ coherent states, the same expression holds except that the last case becomes exactly zero for all $\alpha_0$; it is the superposition states that create oscillations around the recurrence times for finite $\alpha_0$.

We thus find that the effects of quantum coherence, as seen in the correlation functions, depend solely on the difference $\mu-\nu$, which is preserved by operator reordering. In particular, for the symmetrically ordered moments under consideration here, when this difference appropriately aligns with recurrence times we expect strong deviations from the TWA. At other times, the lack of alignment leads to cancellation making phase relationships approximately irrelevant and we expect the TWA to be a good approximation. Furthermore, as $\alpha_0$ increases, quantum deviations from the TWA at times near (but not exactly at) recurrences $T_{\mu \nu}$ become exponentially suppressed as the coherent states in the final case of  Eq.~\eqref{kittenexpectexpanded} become increasingly orthogonal. From our earlier expression for the normally ordered expectation values in the closed system, Eq.~\eqref{normallyorderedexp}, we also see that this suppression is exponential in time, with a dependence of $e^{-|\alpha_0|^2 (t-T_{\mu\nu})^2}$. The quantum deviations from the TWA become more suppressed as $|\alpha_0|$ increases, including the initial short-time deviation  from the TWA \cite{polkovnikov,Polkovnikov2003} and at the  recurrence times, which become increasingly sharp.

We now have the full set of tools needed to consider the expectation value of a moment of a quadrature operator $\langle\hat{X}_\theta^n\rangle$. Any such operator is decomposable into a sum of symmetrically ordered operators 
\bea\label{XThetaDecomp}
\langle \hat{X}_\theta^n \rangle = \sum_{i}^{n} C_{i,n,\theta} \langle \{\hat{a}^{\dagger n-i} \,\hat{a}^{i}\}_{\mathrm{sym}}\rangle,
\eea 
where the index $i$ runs along the even integers for $n$-even, and along the odd integers for $n$-odd. As we established earlier, for every symmetrically ordered expectation value $\langle\{\ha^{\dagger\mu}\ha^\nu\}_{\mathrm{sym}}\rangle $, there is a set of recurrence times $p T_{\mu\nu}$ (with $p\in \mathbb{Z}^+$) where the expectation value is expected to deviate from semiclassical dynamics. For expectation values of the operator $\hat{X}_\theta^n$ given in Eq.~\eqref{XThetaDecomp}, there are a set of times $t^{(n)}_{M,N}$ where at least one of the terms in the expansion of $\hat{X}_\theta^n$ will exhibit recurrence, leading to an overall deviation from the TWA. These times are defined as the recurrence times that satisfy $ \kappa t^{(n)}_{M,N}= 2\pi M/N$ where $M,N\in\mathbb{Z}^+$ and $N$ is even(odd) for $n$ even(odd), and in both cases $N\leq n$. The case with $N=0$ comes from terms with $\mu=\nu$, that is, operators that are proportional to powers of the number operator, and these never exhibit recurrence under the Kerr interaction.\footnote{Note that the number of recurrences between $t=0$ and $t=\frac{2\pi}{\kappa}$ for an operator $\hat{X}_\theta^n$ can be compactly written in terms of the Euler totient summary function, and thus scales quadratically: $\sim \frac{1}{\xi(2)} n^2 + \mathcal{O}(n {\rm log} n)$, where $\xi(2)=\frac{2\pi}{6}$ is the Riemann zeta function evaluated at $2$. In this way, the high spatial frequency features of the state characterized by higher-order moments are connected to high temporal frequency dynamics, exhibiting recurrence more frequently.} Note that the set of recurrence times $t^{(n)}_{M,N}$ for any particular operator $\hat{X}_\theta^n$ is a strict subset of the times $t_{M,N}$ when kitten states form; for the latter $t_{M,N}$, there is no restriction on $M$ and $N$ other than they are positive integers and co-prime (hence, forming a dense set), while for the former $t^{(n)}_{M,N}$, we require $N\leq n$ and $N$ to be even (odd) if $n$ is even (odd).\footnote{We do not require co-prime for the recurrence times as we did for the Kitten times, as we do not need to avoid double counting. Indeed, when there are multiple values of $N$ and $M$ corresponding to the same $t^{(n)}_{M,N}$, we observe a larger deviation from the TWA at the recurrence time due to contributions from multiple terms in the expansion in Eq.~\eqref{XThetaDecomp}.}

\begin{figure*}[t]
\includegraphics[width=1\linewidth]{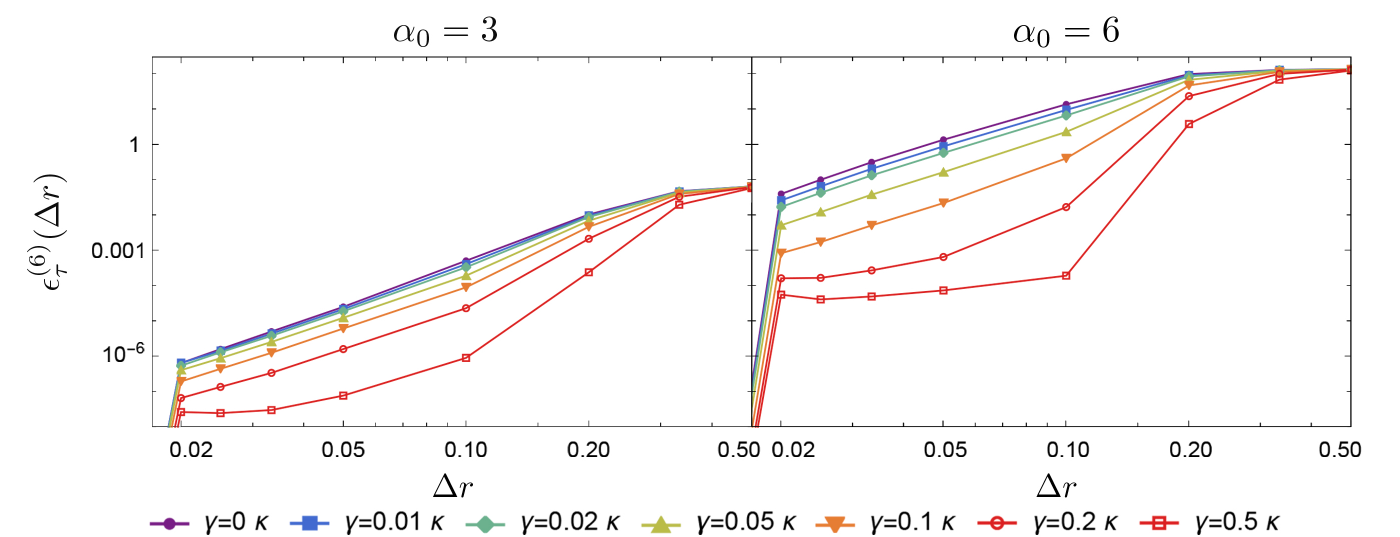} 
     \caption{Error induced by finite grid effects on the calculation of $\langle{\hat{X}_\theta^6}\rangle(t)$ measured relative to $\Delta r_{\rm min} = 10^{-2} \frac{\pi}{\alpha_0^2}$, for $\alpha_0$ varying from $3$ to $6$, and averaging over $20$ values of $\theta$. We define the error in Eq.~\eqref{ErrorDef}, and offset all data by machine precision $10^{-24}$. The maximum time $t_{max}$ is chosen to be $1.3 \frac{\pi}{3\kappa}$ so that, for an initial coherent state at $t=0=t_{\rm min}$, the full behavior of the first recurrence for $\expect{\hat{X}_\theta^6}(t)$ illustrated in Fig.~\ref{PhaseCurves} is captured. We comment that, for $\alpha_0=3$, the ratio of the error induced by discretization for the closed system to the largest $\gamma$ simulation at $\Delta r = 0.1$ is three orders of magnitude larger, whereas for $\alpha_0=6$ the same ratio also at $\Delta r = 0.1$ is four orders of magnitude.}\label{AlphaError}
\end{figure*}

Consider now the deviation of the exact evolution from that given by semiclassical TWA,
\bea\label{DeviationDefinition}
\delta\langle\hat{X}^n_\theta\rangle  (t) \equiv \langle\hat{X}^n_\theta\rangle  (t) - \langle\hat{X}^n_\theta\rangle_{\rm TWA} (t).
\eea 
Whether the deviation from the TWA is maximal  precisely at $t^{(n)}_{M,N}$ depends on the particular phase-quadrature as well as the choice of $\alpha_0$ (due to the fast $\alpha_0$-dependent oscillation in Eq.~\eqref{normallyorderedexp}). We thus define an average deviation from the TWA for our expectation values of interest 
\bea\label{InstantDeviationAverageDef}
\overline{\delta}^{(n)}(t) \equiv \frac{1}{|\alpha_0|^n}\int \frac{d\theta}{2\pi} \bigr|\delta\langle\hat{X}^n_\theta\rangle  (t)|,
\eea 
which we normalize by $|\alpha_0|^n$ so we can meaningfully compare expectation values with differing $\alpha_0$.

As shown in Fig.~\ref{AverageOscillationTime}, at times far from any recurrences $t^{(n)}_{M,N}$ the behavior of the expectation values are well approximated by the TWA (as predicted from the last two cases of Eq.~\eqref{kittenexpectexpanded}). This approximation becomes more exact in the large-$\alpha_0$ limit, as seen with the increasing sharpness of the recurrence  with increasing $\alpha_0$. In this limit the coherent states in the kitten-states become increasingly orthogonal, as detailed above, which also serves to explain the decrease in initial deviations from the TWA at short times as $\alpha_0$ increases, as has been detailed elsewhere \cite{polkovnikov,Polkovnikov2003}. In Fig.~\ref{AverageOscillationTime} we also observe that the TWA improves as $\gamma$ increases, and for the same decoherence rate, the TWA improves for larger $\alpha_0$ as expected in the macroscopic limit. 

Given the average deviation of the exact solution from the TWA (Eq.~\eqref{InstantDeviationAverageDef}), we now define our quantitative metric of interest. We numerically integrate Eq.(\ref{FokkerPlanck}) to find the Wigner function and calculate expectation values based a finite difference method, as described in detail in Appendix~\ref{app:NumericalMethods}. {Given the symmetry, we decompose the Wigner function into a Fourier series in the angular variable and discretize the radial variable with resolution $\Delta r$. The coarseness of this grid determines efficiency of the numerical integration.} Thus, we define a cumulative relative error induced by the phase space discretization in the finite difference method,
\bea\label{ErrorDef}
\epsilon^{(n)}_{ \tau }(\Delta r)\equiv \frac{\int_\tau dt \left |\overline{\delta}^{(n)}_{\Delta r}(t) - \overline{\delta}^{(n)}_{dr}(t) \right |} {\int_\tau dt \left |\overline{\delta}^{(n)}_{dr}(t)\right|}.
\eea 
Here $dr$ is a suitably fine-grained radial grid size that well approximates the exact solution and $\tau$ is a time-window of interest. This is plotted in Fig.~\ref{AlphaError} for the moment $n=6$ with a variety of decoherence rates $\gamma$ and a time window $\tau$ that includes the first recurrence $t^{(6)}_{1,6}=\pi/(3\kappa)$. This metric has the advantage that, if increasing the coarse-grained grid size $\Delta r$ has the effect of over-estimating $\overline{\delta}^{(n)}_{ \tau}$ at some times and under-estimating $\overline{\delta}^{(n)}_{ \tau}$ at other times, these effects  will not cancel out. 

For the cumulative error, in all cases the error resulting from coarse-graining the grid is suppressed for larger decoherence strengths. That this becomes more prominent for larger $\alpha_0$ is made evident by carefully comparing the top and bottom plots in Fig.~\ref{AlphaError}; the separation between the lines corresponding to the closed and open system increases with $\alpha_0$ at intermediate $\Delta r$, especially for the smaller values of $\gamma$ where the subPlanck-scale features are not yet completely washed out. Since these fine-grained features are generated more prominently for larger $\alpha_0$, we expect the error to increase with $\alpha_0$, which we see for the closed system. However, the error induced in calculations of the open system increases more slowly with grid size $\Delta r$. In the limit of asymptotically-strong decoherence, we would expect the error to be completely flat at intermediate $\Delta r$ since no subPlanck features arise.\footnote{At sufficiently high $\Delta r$ there will always be discretization errors due to failure to capture even the semiclassical dynamics given by the TWA, including the open system effect of energy loss.} 

 \subsection{The nature of the quantum-to-classical transition}
From the measurement statistics, we can also understand the nature of the quantum-to-classical transition exhibited by the open Kerr system under photon loss. In contrast to noise models where the quantum signal is merely scaled so the signal-to-noise ratio decreases, such as for a depolarizing channel or a freely decohering cat state, we observe that the concurrent effects of the Kerr nonlinearity together with photon loss leads to qualitatively different open system dynamics. What we will show is that, for the anharmonic oscillator, the quadrature measurement statistics reveal that noise from open system effects do not simply add to the underlying quantum signal but fundamentally change the signal itself.

 To see this, we define a ``trivial'' quantum-to-classical transition to have the form of a convex combination of the closed quantum state and a classical background 
 
\begin{equation}\label{WConvex}
W_{\rm open}(\alpha,t) = p(t) W_{\rm closed}(\alpha,t) + (1-p(t)) W_{\rm classical}(\alpha,t).
\end{equation} Here $W_{\rm open}(\alpha,t)$ is the Wigner function for the exact solution to the full phase space equations of motion for the open quantum system, $W_{\rm closed}(\alpha,t)$ is the evolution given by the closed quantum system\footnote{Although unphysical, a state losing energy without losing coherence also works for $W_{\rm closed}(\alpha,t)$.}, and $p(t)$ is a monotonic function of both time and (implicitly) decoherence strength, starting at $p(0)=1$ and limiting to $p(\infty) = 0$. $W_{\rm classical}(\alpha,t)$ is a classical state in the sense that it lacks quantum coherence and forms a ``background'' for the exact quantum state.  Examples include choosing $W_{\rm classical}(\alpha)$ to be a static fiducial mixed state (e.g. a maximally mixed state as in a depolarizing channel),  a fiducial mixed state evolving under open system evolution (e.g. a mixture of two coherent states for the freely-decohering cat state at short times, Eq.~\eqref{CatWigner}), as we derive in Appendix \ref{app:DecayCat}, or the state of the open quantum system given by the TWA.


Observable expectations naturally inherit the same time dependence,
\begin{equation}\label{OConvex}
    \langle \hat{O}\rangle_{\rm open}(t) = p(t) \langle \hat{O}\rangle_{\rm closed}(t) + (1-p(t)) \langle \hat{O}\rangle_{\rm classical}(t).
\end{equation} Crucially, the difference between the the exact and classical expectation values is simply proportional to $p(t)$, independent of the observable $\hat{O}$ in question:

\begin{equation}\label{DifferenceConvex}\begin{split}
\delta \langle \hat{O}\rangle (t) &\equiv \langle \hat{O}\rangle_{\rm open}(t)-\langle \hat{O}\rangle_{\rm classical}(t)\\
&= p(t)(\langle \hat{O}\rangle_{\rm closed}(t)-\langle \hat{O}\rangle_{\rm classical}(t)).
\end{split}
\end{equation} 

\begin{figure}[t] 
	\includegraphics[width=1\linewidth]{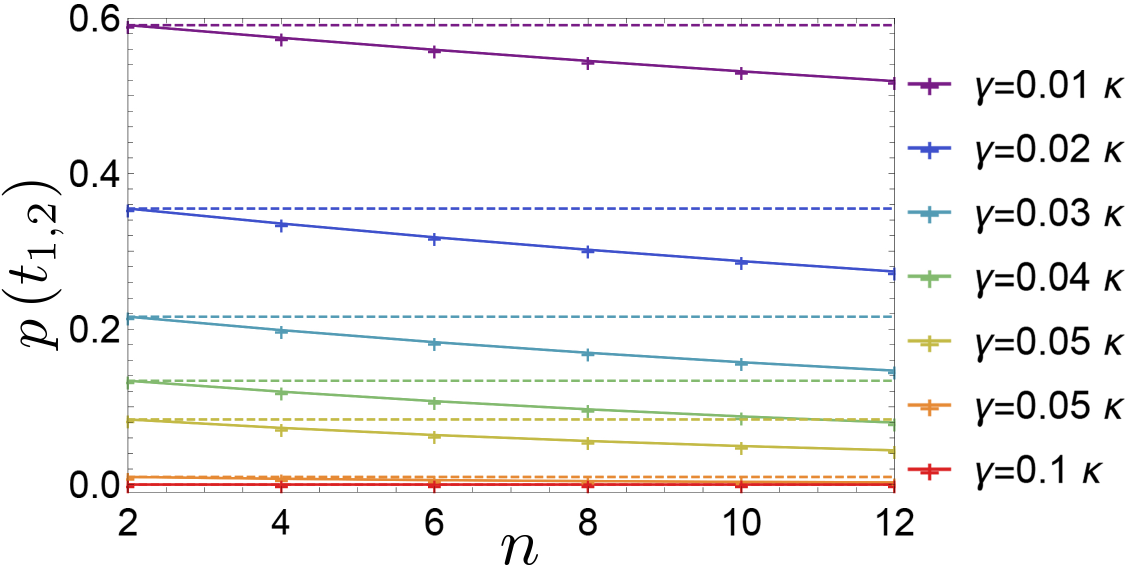} 
	\caption[]{Using the even-powered observables $\langle \hat{X}^n\rangle (t)$, we calculate the value of $p(t)$ at $t=t_{1,2}=\pi/\kappa$, the time of first cat state formation, for which all even-powered observables exhibit recurrence. This is done using numerical data by solving (\ref{DifferenceConvex}) for $p(t)$. If the model indicative of a trivial quantum-to-classical transition defined in (\ref{WConvex}) were accurate, the calculation of $p(t)$ would be independent of $n$ (the horizontal dashed lines), as it would be for a freely decaying cat state or for a depolarizing noise channel. Instead, we see that higher-order observables give rise to a lower calculated value of $p(\pi/\kappa)$, with the effect being most prominent for weaker environmental couplings $\gamma$. Only for the strongest coupling $\gamma = 0.1\kappa$ does the trivial model effectively hold; for $\gamma = 0.1\kappa$, the timescale of decoherence $t_{\rm dec} \sim \frac{1}{\gamma |\alpha_0|^2}$ is sufficiently fast to wipe away all deviations from the TWA by the time of the cat-state formation. } 
	\label{convexitytest}
\end{figure}

For the case where the classical background is the TWA, this is precisely the difference we have studied above. For the depolarizing channel, the classical background of an operator $\hat{O}$ is just its trace, renormalized. For the cases of depolarizing noise and the freely decohering cat state described by Eq.~\eqref{CatWigner}, Eqs.~\eqref{WConvex} and \eqref{OConvex} hold exactly, as it does for any system where the number operator commutes with the Hamiltonian. A consequence is that the decay $\delta \langle \hat{O}\rangle(t)$ (Eq.~\eqref{DifferenceConvex}) is identical for all operators, and thus would be independent of the order of a moment of the quadrature.
For the quadrature moments we have studied, the open system quantum dynamics are indeed fluctuations on top of the semiclassical behavior given by the TWA, which is consistent with Eq.~\eqref{OConvex}. However, when we compare different orders of the moments for the open system, the fluctuations are suppressed at different rates.  This is shown in Fig.~\ref{convexitytest} where we calculate the function $p(t)$ for different expectation values $\langle \hat{X}^n \rangle(t)$, and evaluate it at $t=\pi/\kappa$ (the time of cat state formation). Here, we see that the function $p(t)$ decays faster for higher-order moments, so that the $\delta \langle \hat{O}\rangle$ also decays faster for expectation values of higher-order operators.\footnote{For qubits, the analogy of these higher-order quadrature moments are the difference of  multi-qubit observables from their TWA-values.} This means that the function $p(t)$ in Eq.\eqref{OConvex} \emph{must} depend on the observable $\hat{O}$. 

For the open Kerr system, the quantum-to-classical transition is not a trivial convex noise model, as in a depolarizing channel often used in studying NISQ devices, but is instead a situation where the closed quantum effects interact with the open system effect to give rise to genuinely new behavior. Opening the quantum system does not just add noise to a quantum signal, it changes the nature of the signal itself. 
In this case, the open system is not a simulation of the closed system with noise added on top. It is its own system with entirely different dynamics.

\section{Summary and Outlook}
In this work we have studied the effect of decoherence on the efficiency of simulating quantum dynamics in the case of an anharmonic Kerr oscillator, a standard paradigm in quantum optics.  As the decoherence facilitates the quantum-to-classical transition and washes out quantum complexity, we expect more efficient representations are possible for the open quantum system.  To quantify this, we studied quantum phase space dynamics represented by the Wigner function.  Whereas the closed-system quantum dynamics leads to fine-grained subPlanck scale structure, decoherence acts to wash this out, making the semiclassical description according to the TWA more accurate.  This semiclassical dynamics is more efficiently simulated by a coarse-grained finite-difference integration of the Wigner function, when including the Fokker-Planck terms associated with decoherence. 

For the specific cases of the Kerr oscillator we showed how semiclassical and quantum dynamics diverge due to quantum interference and the generation of superpositions of coherent states.  Without the fine structure refeeding coherence to the system, evolution of the open system does not produce these high-order kitten states. The result of this is a reduction of the quantum deviations from the TWA (Fig.~\ref{AverageOscillationTime}) and a suppression of numerical errors induced by simulation of the system on a coarser discrete grid for numerical integration (Fig.~\ref{AlphaError}). 

The tendency towards expectation value dynamics given by the TWA is well-known in the literature \cite{Polkovnikov2003}. Here, we have explained this in terms of the breaking of the $\mathbb{Z}_n$ symmetry of the kitten states by decoherence concurrent with Hamiltonian evolution. We have also shown that this reduction in computational cost is not a reduction in signal-to-noise ratio (such as is the case for the depolarizing noise model for qubits), but rather a nontrivial transition from quantum dynamics to classical dynamics.  For the anharmonic oscillator at zero temperature, the open system is not a noisy simulation of the closed system but a different (and easier) problem entirely, due to decoherence preventing the formation of fine-scale structure in the first place. To show this, we have also introduced a new test of the structure of the quantum-to-classical transition based on the sensitivity of different order observables to decoherence. In future work we plan to extend this test to other systems, e.g., collections of qubits, to better guide the use of appropriate noise models.

Our work makes quantitative the intuition long established in the early work on decoherence. Decoherence reduces the complexity of the quantum state in phase space, and it does so more prominently for more macroscopic initial states.  This translates into more efficient simulation of the open quantum system as the system size grows. While we have studied this for the toy problem of a single anharmonic oscillator, its implications for obtaining a quantum advantage in NISQ devices is another open question. How does the rapidity of the regression to semiclassical dynamics with increasing system size generalize for mulitpartite systems such as multiple coupled nonlinear oscillators or qubits, and what is an efficient representation of the open quantum systems that would allow for efficient classical simulations?  

The current work gives some hints in this direction.  Phase space representations may provide for scalable efficient simulations of open quantum systems when decoherence is sufficiently large. Coarse graining and well-chosen finite discretization, informed by semiclassical dynamics is one potential method, as seen in recent work~\cite{Llordes2023}.  Another potential method is mapping the phase space dynamics to an underlying set of stochastic Langevin equations.  While such a method cannot efficiently capture all complex quantum dynamics, it may do so in the presence of decoherence~\cite{Piotr2021}. It remains an open question whether, despite the nontrivial quantum-to-classical transition,\footnote{For a trivial quantum-to-classical transition, one can always obtain statistics of the closed system at time $t$ by sampling the noisy open system $\sim1/p(t)$ times.} the complexity of simulation of the open and closed systems are related by a polynomial overhead i.e. via zero-noise extrapolation~\cite{GiurgicaTiron2020}. In future work we will study this to better understand when the quantum-to-classical transition is also a transition in computational complexity. 
\\
\section{Acknowledgements} 

We gratefully acknowledges helpful conversations with Jun Takahashi, Changhao Yi, and Chris Jackson. 

The authors would like to thank the UNM Center for Advanced Research Computing, supported in part by the National Science Foundation, for providing the high performance computing resources used in this work. 
This work was supported by National Science Foundation Grant No. PHY-2116246 and Grant No. 2037755, and is based upon work partially supported by the U.S. Department of Energy, Office of Science, National Quantum Information Science Research Centers, Quantum Systems Accelerator. 
This material is based upon work supported by the Air Force Office of Scientific Research under award number FA9550-22-1-0498. Any opinions, findings, and conclusions or recommendations expressed in this material are those of the author(s) and do not necessarily reflect the views of the United States Air Force.

We acknowledge the indigenous peoples of the Pueblo of Sandia as the original inhabitants, stewards, and protectors of the lands on which the University of New Mexico now sits. 

\bibliography{Kerr_Decoherence}
\newpage

\appendix

\section{Summary of Numerical Approach} \label{app:NumericalMethods}
Our aim is to numerically solve the (linear) partial differential equation (PDE) for the Wigner function in a dissipative medium with a self-Kerr interaction in cylindrical coordinates~\cite{stobinska2008}
\begin{widetext}
\bea \label{eqt:pde1}
    \partial_\tau W(\tau, r, \varphi) &=& \left\{ \underbrace{(r^2 - 1) \partial_\varphi}_{\text{mean field rotation}} - \frac{1}{16} \left( \underbrace{\frac{1}{r} \partial_r}_{\text{Gaussian shearing}}  + \underbrace{\partial_r^2 + \frac{1}{r^2} \partial_{\varphi}^2}_{\text{non-Gaussian rotation}} \right) \partial_\varphi \right. \nonumber  \\
&& \left. + \underbrace{\xi + \frac{\xi}{2} r  \partial_r}_{\text{drift}} + \underbrace{\frac{\xi}{8} \left( \partial_r^2 + \frac{2}{r} \partial_r +\frac{1}{r^2} \partial_\varphi^2 \right)}_{\text{diffusion}} \right\} W(\tau, r, \varphi) \ ,
\eea
\end{widetext}
where we have defined dimensionless time $\tau = \kappa t$ and dimensionless decay $\xi = \frac{\gamma}{\kappa}$ with boundary conditions
%


\beq
W(\tau, 0, 0) = \frac{2}{\pi} e^{-2 |\alpha_0|^2 e^{-\tau \xi}} \ , \quad \lim_{r \to \infty} W(\tau, r, \varphi) = 0 \ ,
\eeq 
and initial condition
\beq \label{eqt:IC1}
W(0, r, \varphi) = \frac{2}{\pi} e^{-2 |\alpha_0 - r e^{i \varphi} |^2} \ .
\eeq
Without loss of generality assuming $\alpha_0$  to be real, we have
\begin{eqnarray}
W(0, r, \varphi) &=& \frac{2}{\pi} e^{-2 (\alpha_0^2 + r^2)} e^{4 \alpha r \cos \varphi} \nonumber \\ 
& & \hspace{-2cm} = \frac{2}{\pi} e^{-2 (\alpha_0^2 + r^2)} \left[ I_0(4 r \alpha_0) + 2 \sum_{k=1}^\infty I_k(4 r \alpha) \cos(k \varphi)\right] \nonumber \ , \\
\end{eqnarray}
where $I_k(x)$ are the modified Bessel functions of the first kind. Because the coefficients of Eq.~\eqref{eqt:pde1} are independent of $\varphi$ and the Wigner function must be periodic in $\varphi$, we choose to expand the Wigner function in a sine/cosine series
\begin{eqnarray}
W(\tau, r, \varphi) &=& a_0 (\tau, r) + \sum_{k=1}^{\infty} \left( a_k(\tau,r) \cos (k \varphi)  \right. \nonumber \\
&& \left. + b_k(\tau,r) \sin (k \varphi) \right) \ .
\end{eqnarray}
The normalization condition can now be expressed as
\beq
\int_0^{2\pi} d \varphi \int_{0}^{\infty} dr \ r W(\tau, r, \varphi) = 2 \pi \int dr \ r a_0(\tau, r)= 1 \ .
\eeq

Plugging the expansion of the Wigner function into Eq.~\eqref{eqt:pde1}, we get a coupled system of ordinary differential equations (ODE's) to solve for each $k$ value
\begin{widetext}
\bea \label{eqt:pde2}
\partial_\tau a_0(\tau, r) &=& \left\{ \xi + \frac{\xi}{2} \left( r + \frac{1}{4r} \right) \partial_r + \frac{\xi}{8}   \partial_r^2  \right\} a_0(\tau, r) \ , \\
\partial_\tau a_k(\tau, r) &=& \left\{ k (r^2 - 1) - \frac{k}{16} \left( \frac{1}{r} \partial_r  + \partial_r^2 - \frac{k^2}{r^2} \right) \right\} b_k(\tau, r)  \nonumber  \\
&& + \left\{ \xi + \frac{\xi}{2} \left( r +  \frac{1}{4r} \right) \partial_r + \frac{\xi}{8} \left( \partial_r^2 - \frac{k^2}{r^2}  \right) \right\} a_k(\tau, r)  \ , \\
\partial_\tau b_k(\tau, r) &=& \left\{ -k (r^2 - 1) + \frac{k}{16} \left( \frac{1}{r} \partial_r  + \partial_r^2 - \frac{k^2}{r^2} \right) \right\} a_k(\tau, r)  \nonumber  \\
&& + \left\{ \xi + \frac{\xi}{2} \left( r + \frac{1}{4r} \right) \partial_r + \frac{\xi}{8} \left( \partial_r^2 - \frac{k^2}{r^2}  \right) \right\} b_k(\tau, r) \ .
\eea
\end{widetext}
The equations for different $k$'s do not couple, but for each $k$ the pair $a_k(\tau,r)$ and $b_k(\tau,r)$ are governed by coupled ODE's. The boundary condition at $r = 0$ is given by
\beq
a_0(\tau,0) = \frac{2}{\pi} e^{-2 |\alpha_0|^2 e^{-\tau \xi}} \ , \quad a_k(\tau, 0) = 0 \ , \quad b_k (\tau, 0 ) = 0 \ ,
\eeq
and the boundary condition at $r \to \infty$ is given by
\bes
\begin{align}
\lim_{r \to \infty} a_0(\tau,r) & = 0 \ , \\
\lim_{r \to \infty} a_k(\tau,r) & = 0 \ , \quad k>0 \ , \\
\lim_{r \to \infty} b_k(\tau,r) & = 0 \ , \quad k>0 \ ,
\end{align}
\ees
The initial condition is given by (again, assuming $\alpha_0$ is real)
\bes \label{eqt:IC2}
\begin{align}
a_0(0,r) & = \frac{1}{\pi^2} e^{-2 (\alpha_0^2 + r^2)} \int_{-\pi}^{\pi} d \varphi \  e^{4 r \alpha_0 \cos(\varphi)} \nonumber \\
& = \frac{2}{\pi} e^{-2(\alpha_0^2 + r^2)} I_{0}(4 r \alpha_0) \ , \\
a_k(0,r) & = \frac{2}{\pi^2} e^{-2 (\alpha_0^2 + r^2)} \int_{-\pi}^{\pi} d \varphi \ e^{4 r \alpha_0 \cos(\varphi)} \cos (k \varphi) \nonumber \\& = \frac{4}{\pi} e^{-2(\alpha_0^2 + r^2)} I_{k}(4 r \alpha_0) \ , \quad k > 0 \ , \\
b_k(0,r) & = 0 \ , \quad k>0 \ .
\end{align}
\ees
From the initial condition, we can find the leading behavior for the functions near $r = 0$
\bes
\begin{align}
a_0(0, r) & = \tilde{a}_0 + r^2 \tilde{\tilde{a}}_0 + \dots \ , \\
a_k(0, r) & = \tilde{a}_k r^k + \tilde{\tilde{a}}_k r^{k+2} + \dots \ .
\end{align}
\ees
We find that this provides a consistent expansion for the functions at $r = 0$ even for $\tau > 0$
\bes \label{eqt:BC}
\begin{align}
a_0(\tau, r) & = \tilde{a}_0(\tau)+ r^2 \tilde{\tilde{a}}_0(\tau) + \dots \ , \\
a_k(\tau, r) & = \tilde{a}_k(\tau) r^k + \tilde{\tilde{a}}_k r^{k+2}(\tau) + \dots \ , \\
b_k(\tau, r) & = \tilde{b}_k(\tau) r^k + \tilde{\tilde{b}}_k r^{k+2}(\tau) + \dots \ . 
\end{align}
\ees

The ODE's are solved by discretizing the radial direction  into $n$ equal steps
\beq
r_j = j \Delta r \ , \quad j = 1, \dots n \ ,
\eeq
with $\Delta r = r_{\mathrm{max}}/n$, where $r_{\mathrm{max}}$ being some sufficiently large $r$ value.  In our simulations, we take $r_{\mathrm{max}} = 2 \alpha_0$. We will use the index $j = 0$ to denote the boundary condition at $r = 0$.  The derivatives discretized up to fourth order are given by
\begin{widetext}
\bea
\partial_r f(\tau,r) &\rightarrow& \frac{ f_{j-2}(\tau) -8 f_{j-1}(\tau) + 8 f_{j+1}(\tau) - f_{j+2}(\tau)}{12 \Delta r} \ , \\
\partial_r^2 f(\tau,r) &\rightarrow& \frac{ -f_{j-2}(\tau) +16 f_{j-1}(\tau) -30 f_j(\tau) + 16 f_{j+1}(\tau) - f_{j+2}(\tau)}{12 \Delta r^2} \ ,
\eea
\end{widetext}
where $f_j(\tau) \equiv f(\tau, r_j)$.  Let us define the vector $\vec{g}_k(\tau) = \left(a_{k,1}(\tau),  \dots, a_{k,n}(\tau), b_{k,1}(\tau), \dots b_{k,n}(\tau)\right)^T$.  We will then write our ODE's as a system of linear equations to be solved
\bes \label{eqt:linearSystem}
\begin{align}
\partial_t \vec{g}_0(\tau) &= A_0 \vec{g}_0(\tau) + \vec{b}_0(\tau) \ , \\
\partial_t \vec{g}_k(\tau) &= A_k \vec{g}_k(\tau) \ , k > 0 \ . 
\end{align}
\ees
The vector $\vec{b}_0(\tau)$ arises from the discretization near $r = 0$: when terms such as $a_{0,0}(\tau)$ appear when evaluating the ODE at $j = 1$ and $j=2$, we must replace them with the initial condition
\beq
a_{0,0}(\tau) = \frac{2}{\pi} e^{-2 \alpha^2} \ ,
\eeq
which then represent constants in the system of equations. (Recall that $a_{k,0}(\tau) = 0$ so this term does not appear for $k>0$.)
The terms $a_{k,-1}(\tau)$ and $b_{k,-1}(\tau)$ that arise at $j=1$ must be replaced by
\beq
a_{k,-1}(\tau) = (-1)^k a_{k,1}(\tau) \ , \quad b_{k,-1}(\tau)= (-1)^k b_{k,1}(\tau) \ .
\eeq
This choice is to enforce the even/odd behavior from Eq.~\eqref{eqt:BC}.  We note that this term only appears from the term $\frac{\xi}{2} r \partial_r$, and it is cancelled otherwise. We enforce the boundary condition at infinity by setting to zero terms beyond $r_{\mathrm{max}}$, i.e. $a_{k,n+1}(\tau) = a_{k,n+2}(\tau) = b_{k,n+1}(\tau) = b_{k,n+2}(\tau) = 0$.  

To solve the system of linear equations in Eq.~\eqref{eqt:linearSystem}, we use TR-BDF2 \cite{TRBDF2}, an implicit Runge-Kutta method. We use an absolute error tolerance of $10^{-9}$ and a relative error tolerance of $10^{-6}$.  Our maximum temporal step size is chosen to be $\pi/(100 \alpha_0^2)$.  We simulate pairs of $(a_k,b_k)$ up to a maximum $k$ value of 60.

After solving for the time-evolved Wigner function up to some time $\tau$, we can plot the Wigner function in terms of $(x,p)$ for fixed $\kappa t$ as was done in the main text by plotting the function $a_0(\tau,r) + \sum_{k=1}^{\infty} \left(a_k(\tau,r) \cos(k \varphi) + b_k(\tau,r) \sin(k \varphi) \right)$.  Our $\varphi$ grid is discretized in steps of $\pi/200$.

We can also calculate the expectation value of an observable $\hat{A}$ as (we drop the time dependence for now)
\beq
\langle \hat{A} \rangle = \int_{-\infty}^{\infty} d x \int_{-\infty}^{\infty} dp \  W(x,p) \tilde{A}(x,p) \ ,
\eeq
where $\tilde{A}(x,p)$ is the Weyl transform of the operator $\tilde{A}$
\begin{eqnarray}
\tilde{A}(x,p) &=& \int d y e^{-i p y /\hbar} \langle x + \frac{y}{2} | \tilde{A} | x - \frac{y}{2} \rangle \nonumber \\
&=& \int d u e^{i x u / \hbar} \langle p + \frac{u}{2} | \tilde{A} | p - \frac{u}{2} \rangle \ .
\end{eqnarray}
In particular, the expectation value of arbitrary powers of $\hat{X}$ and $\hat{P}$ is given by
\beq
 \langle \hat{X}^m \hat{P}^n \rangle =  \int_{-\infty}^{\infty} d x \int_{-\infty}^{\infty} dp \  x^m p^n W(x,p) \ ,
\eeq
Using our solution for the Wigner function in polar coordinates, this takes the form
\begin{widetext}
\bea
\langle \hat{X}^m \hat{P}^n \rangle &=&  \sum_{k=0}^{\infty} \left[ \int_{0}^{2 \pi} d \varphi \ \cos^m \varphi \sin^n \varphi \cos(k \varphi) \int_{0}^{\infty} dr \  r^{m+n+1} a_k(r)  \right. \nonumber \\
&& \left. + \int_{0}^{2 \pi} d \varphi \ \cos^m \varphi \sin^n \varphi \sin(k \varphi) \int_{0}^{\infty} dr \  r^{m+n+1} b_k(r)  \right] \ .
\eea
\end{widetext}
The integral over $\varphi$ can be calculated (numerically) by noting that
\begin{widetext}
\bes
\begin{align}
\int_{0}^{2 \pi} d \varphi \ \cos^m \varphi \sin^n \varphi \cos(k \varphi) &= \frac{2 \pi}{2^{m+n+1}i^n} \sum_{k_m=0}^m \sum_{k_n=0}^n \binom{m}{k_m} \binom{n}{k_n} (-1)^{k_n} \left( \delta_{m+n+k- 2k_m -2k_n,0} + \delta_{m+n-k- 2k_m -2k_n,0} \right) \ , \\
\int_{0}^{2 \pi} d \varphi \ \cos^m \varphi \sin^n \varphi \sin(k \varphi) &= \frac{2 \pi}{2^{m+n+1}i^{n+1}} \sum_{k_m=0}^m \sum_{k_n=0}^n \binom{m}{k_m} \binom{n}{k_n} (-1)^{k_n} \left( \delta_{m+n+k- 2k_m -2k_n,0} - \delta_{m+n-k- 2k_m -2k_n,0} \right) \ .
\end{align}
\ees
\end{widetext}
In this form, it is clear that we need only to perform radial integrals of the form $\int_{0}^{\infty} dr \  r^{p+1} a_k(r), b_k(r)$. 

\section{Symmetrically ordered expectation values from normally ordered ones}\label{app:symtonorm}
Let the label $s=\{-1,0,+1\}$ denote anti-normal ordering, symmetric ordering, and normal ordering, respectively. Here we derive the relationship \eqref{expectschars} connecting symmetric and normal ordered correlation functions.

One has $\langle \{\hat{a}^{\dag n}\hat{a}^{m}\}_s\rangle=\int d^2\alpha W_{-s}(\alpha,\alpha^*)(\alpha^{*})^n(\alpha)^m$, where $W_s(\alpha,\alpha^*)$ is the $s$-ordered quasiprobability function; $W_0$ is the Wigner function, $W_{+1}$ is the Husimi Q function, and $W_{-1}$ is the Glauber-Sudarshan P function. An $s$-ordered correlation is directly computed by the appropriate integral over the $(-s)$-ordered quasiprobability function; for example, integrating against the P function gives normally ordered correlation functions.

The $s$-ordered characteristic function is the (2d) Fourier transform of the corresponding quasiprobability function,
\begin{equation}
    \chi_s (\beta,\beta^*)=\int \frac{d^2\alpha}{\pi}W_s(\alpha,\alpha^*)e^{\beta \alpha^*-\beta^*\alpha}.
\end{equation}
It follows that the characteristic function is the moment generating function
\begin{equation}
    \frac{\partial^n}{\partial \beta^n}\frac{\partial^m}{\partial (-\beta^*)^m}\chi_s(\beta,\beta^*)\biggr|_{\beta=0}=\frac{1}{\pi}\langle \{\hat{a}^{\dag n}\hat{a}^{m}\}_{-s}\rangle.
\end{equation}
One can also observe that $\chi_s(\beta,\beta^*)=\text{tr}(\rho D_{-s}(\beta,\beta^*))$ where $D_s(\beta,\beta^*)=e^{\frac{s|\beta|^2}{2}}e^{\beta \hat{a}^\dag-\beta^*\hat{a}}$, and so, in particular, $\chi_0(\beta,\beta^*)=e^{\frac{-|\beta|^2}{2}}\chi_{-1}(\beta,\beta^*)$. Thus, we have
\begin{equation}\label{notdistributed}
   \frac{1}{\pi}\langle \{\hat{a}^{\dag n}\hat{a}^{m}\}_0\rangle =(-1)^m\frac{\partial^n}{\partial \beta^n}\frac{\partial^m}{\partial \beta^{*m}}\left(e^{\frac{-|\beta|^2}{2}}\chi_{-1}(\beta,\beta^*)\right)\biggr|_{\beta=0}.
\end{equation}

\begin{figure}[t] 
	\includegraphics[width=.9\linewidth]{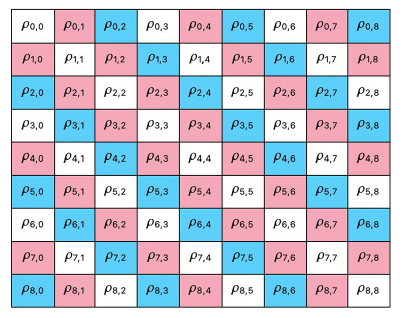} 
	\caption[]{For open quantum systems with anharmonic, number-preserving Hamiltonians, each diagonal stripe of the density matrix in the number basis (colored) evolves independently. }
	\label{densityColors}

\end{figure}

We will obtain a relation between the symmetric and normally ordered correlations by distributing the derivatives. 
\begin{widetext}
First note
\begin{equation}
    \frac{\partial^n}{\partial x^n}(e^\frac{-xy}{2}f(x,y))=e^\frac{-xy}{2}\left(\frac{-y}{2}+\frac{\partial}{\partial x}\right)^nf(x,y)
    =\sum_{\nu=0}^n\binom{n}{\nu}\left(\frac{-y}{2}\right)^\nu e^\frac{-xy}{2}\frac{\partial^{n-\nu}}{\partial x^{n-\nu}}f(x,y),
\end{equation}
and thus we have
\begin{equation}\label{firstderivdistribution}
        \frac{\partial^m}{\partial y^m}\frac{\partial^n}{\partial x^n}(e^\frac{-xy}{2}f(x,y))=\sum_{\nu=0}^n \binom{n}{\nu} \frac{\partial^m}{\partial y^m}(e^\frac{-xy}{2}g(x,y))
        =\sum_{\nu=0}^n\sum_{\mu=0}^m\binom{n}{\nu}\binom{m}{\mu}e^\frac{-xy}{2}\left(\frac{-x}{2}\right)^\mu\frac{\partial^{m-\mu}}{\partial y^{m-\mu}}g(x,y),
\end{equation}
where 
\begin{equation}
    g(x,y)=\left(\frac{-y}{2}\right)^\nu\frac{\partial^{n-\nu}}{\partial x^{n-\nu}}f(x,y).
\end{equation}
Now, observing that 
\begin{equation}
    \frac{d^k}{dx^k}\left((ax)^l h(x)\right)=\sum_{p=0}^k\binom{k}{p}\frac{l!}{(l-p)!}a^lx^{l-p}\frac{d^{k-p}}{dx^{k-p}}h(x),
\end{equation}
we can write

\begin{equation}
\frac{\partial^{m-\mu}}{\partial y^{m-\mu}}g(x,y)
    =\sum_{\gamma=0}^{m-\mu}\binom{m-\mu}{\gamma}\frac{\nu!}{(\nu-\gamma)!}\left(\frac{-1}{2}\right)^\nu y^{\nu-\gamma}\frac{\partial^{m-\mu-\gamma}}{\partial y^{m-\mu-\gamma}}\frac{\partial^{n-\nu}}{\partial x^{n-\nu}}f(x,y)
\end{equation}
which, when substituted into \eqref{firstderivdistribution} gives
\begin{equation}
   \frac{\partial^m}{\partial y^m}\frac{\partial^n}{\partial x^n}(e^\frac{-xy}{2}f(x,y))
        =\sum_{\nu=0}^{n}\sum_{\mu=0}^m\sum_{\gamma=0}^{m-\mu}\binom{n}{\nu}\binom{m}{\mu}\binom{m-\mu}{\gamma}\frac{\nu!}{(\nu-\gamma)!}\left(\frac{-1}{2}\right)^{\mu+\nu}e^\frac{-xy}{2}x^\mu y^{\nu-\gamma}\frac{\partial^{m-\mu-\gamma}}{\partial y^{m-\mu-\gamma}}\frac{\partial^{n-\nu}}{\partial x^{n-\nu}}f(x,y).
\end{equation}
Evaluating this at $x=y=0$ results in setting $\mu=0$ and $\nu=\gamma$ as only these terms avoid explicit $x$ and $y$ dependence. Hence, after replacing combinatorial terms with factorials and simplifying,
\begin{equation}\label{generalderivdistribute}
    \frac{\partial^m}{\partial y^m}\frac{\partial^n}{\partial x^n}(e^\frac{-xy}{2}f(x,y))\biggr|_{x=y=0}=\sum_{\gamma=0}^{\text{min}(m,n)}\frac{n!m!}{\gamma!(n-\gamma)!(m-\gamma)!}\left(\frac{-1}{2}\right)^\gamma\frac{\partial^{m-\gamma}}{\partial y^{m-\gamma}}\frac{\partial^{n-\gamma}}{\partial x^{n-\gamma}}f(x,y)\biggr|_{x=y=0}.
\end{equation}
With this, we can express symmetrically ordered correlation functions in terms of normally ordered ones by applying \eqref{generalderivdistribute} to
\eqref{notdistributed},
\begin{equation}
\begin{split}
    \langle \{\hat{a}^{\dag n}\hat{a}^{m}\}_0\rangle&=\pi(-1)^m\sum_{\gamma=0}^{\text{min}(m,n)}\frac{n!m!}{\gamma!(n-\gamma)!(m-\gamma)!}\left(\frac{-1}{2}\right)^\gamma\frac{\partial^{n-\gamma}}{\partial \beta^{n-\gamma}}\frac{\partial^{m-\gamma}}{\partial (\beta^*)^{m-\gamma}}\chi_{-1}(\beta,\beta^*)\biggr|_{\beta=\beta^*=0}\\
    &=\sum_{\gamma=0}^{\text{min}(m,n)}\frac{n!m!}{\gamma!(n-\gamma)!(m-\gamma)!}\left(\frac{1}{2}\right)^\gamma\langle \{\hat{a}^{\dag n-\gamma}\hat{a}^{m-\gamma}\}_{+1}\rangle,
    \end{split}
\end{equation}
which we recognize as relation \eqref{expectschars}.

\section{Derivation of Recurrence Time Formulae and Zero-Temperature Fock-Space Solution for the Open Quantum System}\label{app:reccurencethms}

In Eq.~\eqref{densityDiscreteFourier}, we observe that the density matrix for a bosonic mode with an anharmonic, number-preserving Hamiltonian has the formal solution 
\bea\label{densityDiscreteFourierRepeat}
\rho_{jk}(t) = \sum_{j',k'} A_{j'k'}^{(j,k)} e^{-\tilde{\gamma}_{j'k'} t} \delta_{j'-k', j-k} ,
\eea
where $\tilde{\gamma}_{j'k'}$ is a complex coefficient specific to each density matrix element, depending on both the Hamiltonian operator and Lindbladian superoperator's elements in the Fock basis. Such an equation is a solution to the family of equations 

\bea\label{coupledequations}
\frac{d}{dt} \rho_{jk}(t) = - \tilde{\gamma}_{jk} \rho_{jk}(t) + \sum_n \left(g^{(n)\,+}_{jk} \rho_{j+n,k+n}(t) + g^{(n)\,-}_{jk} \rho_{j-n,k-n}(t) \right) ,
\eea
which are generated by the Lindblad master equation with multiple arbitrary $n$-photon-loss/gain channels, described by a coupling $g^{(n)\,\pm}_{jk}$ respectively. Recast in this way, Eq.~\eqref{coupledequations} has the form of a set of coupled systems where population can leak out either to other systems or to an environment.\footnote{Explicitly, we work with a time domain generalization of Eq.~(28) in \cite{ProppNet}, making a reindexing substitution $\rho_{jk}\rightarrow c_i$, which recasts the evolution of a single stripe of the density matrix, as depicted in Fig.~\ref{densityColors}, with respect to an anharmonic Hamiltonian and open system effects as energy moving between simple harmonic oscillators which are coupled to each other and to an unmonitored environment.} Using techniques for such systems \cite{ProppNet}, we can solve each diagonal stripe of the density matrix separately, as depicted in Fig.~\ref{densityColors} for a small number of basis elements. \end{widetext}

If we specialize to the Kerr Hamiltonian defined in Eq.~\eqref{Hamiltonian} and zero-temperature with amplitude damping (that is, single photon loss only) giving rise to the Lindbladian in Eq.~\eqref{ampdampLindbladian}, the complex decay rates $\tilde{\gamma}_{jk}$ have the form 
\bea\label{gammajkCoeffs3}
\tilde{\gamma}_{jk} = \frac{\gamma}{2}(j+k) + i\frac{\kappa}{2}\left( (j^2 -j) - (k^2-k)\right) ,
\eea 
and the couplings are given by
\bea\label{couplingdef}
g^{(1)\,+}_{jk} = \gamma \sqrt{(j+1)(k+1)} ,
\eea 
with all other $g^{(n)\,\pm}_{jk} = 0$.

Returning to the more general case of an anharmonic number-preserving oscillator with arbitrary $n$-photon loss and gain channels, we will now show that the two results derived in the main text hold under this more general case: recurrence times are independent of both operator ordering and open system effects. Here we define $s=+1$ to be normally ordered, $s=0$ to be symmetrically ordered, and $s=-1$ to be anti-normally ordered. 

We can decompose the expectation value of an $s$-ordered product of powers of the creation and annihilation operators in terms of a single stripe of the density matrix
\bea\label{expectsDirectv2}
\expect{\hadmu \ha^\nu}_s (t) =\sum\limits_{jk} P_{jk}^{\mu\nu s} \rho_{jk}(t) \delta_{j-k,\mu-\nu} ,
\eea 
with $P_{jk}^{(\mu \nu s)}$ a positive and real combinatorial factor. Substituting in the solution for density matrix elements given in Eq.~\eqref{densityDiscreteFourierRepeat}, we find 
\bea\label{expectDiscreteFourier2}
\expect{\hadmu \ha^\nu}_s (t) = \sum_{j'k'}
R_{j'k'}^{\mu\nu s} e^{-\tilde{\gamma}_{j'k'} t}  \delta_{j'-k',\mu-\nu} ,
\eea 
where we have defined new Fourier coefficients for the expectation value 
\bea\label{FourierCoeffs2}
R_{j'k'}^{\mu\nu s} = \sum_j P_{j'-k'+j,j}^{\mu\nu s} A_{j'k'}^{j'-k'+j,j}.
\eea 
where we reindexed $j\rightarrow j'-k'+j$ and contracted the Kronecker product.

Note that it is the sums defining the Fourier coefficients in Eq.~\eqref{FourierCoeffs2} that collapse to a single term for the closed system, and not the sum in the expectation value in Eq.~\eqref{expectDiscreteFourier2}. Thus, there is a set of frequencies $\omega_j$ present in the time-dependent expectation value, where $\omega_j = {\rm Im}[\tilde{\gamma}_{j,j+\mu-\nu}] $ for $\mu>\nu$ and $\omega_j = {\rm Im}[\tilde{\gamma}_{j+\nu-\mu,j}] $ for $\mu<\nu$ (the two cases ensure that the indices on $\tilde{\gamma}_{j,k}$ remain nonnegative). For any anharmonic number-preserving Hamiltonian, all these frequencies are distinct. 

To see this, consider a Hamiltonian decomposed into the form
\bea\label{AnharmonicH}
H_{\rm anh} = \sum_k b_k \hat{n}^k .
\eea 
We see that, for an expectation value $\expect{\hadmu \ha^\nu}_s (t)$ (assuming $\mu<\nu$) the frequencies for the closed system will have the form 
\bea\label{omegas}
\omega_j = \sum_k b_k (j^k  - (j+\nu-\mu)^k),
\eea 
which are negative in this case. For the harmonic case, $b_1$ is the only non-zero term, and the frequencies $\omega_j $ within a stripe are independent of $j$ and thus identical (or when $\mu=\nu$ so that there simply is no frequency). In all other cases, each frequency depends on $j$ and thus the sum in Eq.~\eqref{expectDiscreteFourier2} will not result in cancellations of any particular frequency, no matter what the Fourier coefficients $R_{j'k'}^{\nu\mu s}$ may be, provided they are nonzero.

That they are nonzero for the closed system is trivial; the sum $R_{j'k'}^{\nu\mu s}$ reduces to a single term corresponding to the initial conditions for a density matrix element $\rho_{j'k'}(0)$ multiplied by a positive combinatorial factor. 

To show that the Fourier coefficients are non-zero for the open system is the final ingredient to complete our more rigorous argument. We begin by noting that the effect of arbitrary $n$-photon gain and loss channels is to modify the real part of $\tilde{\gamma}_{jk}$ as well as the coefficients $A_{j'k'}^{j'-k'+j,j}$. 

This latter point is seen by inserting the formal solution for the density matrix elements Eq.~\eqref{densityDiscreteFourierRepeat} into the coupled equations Eq.~\eqref{coupledequations},
\begin{widetext}
\bea\label{betterarg2}
\frac{d}{dt}\rho_{jk}(t)&=& -\sum_{j'k'} \tilde{\gamma}_{j'k'} A_{j'k'}^{(jk)} e^{-\tilde{\gamma}_{j'k'}t}\delta_{j'-k',j-k} \\
&=& -\tilde{\gamma}_{jk} \sum_{j'k'} A_{j'k'}^{(jk)} e^{-\tilde{\gamma}_{j'k'}t}\delta_{j'-k',j-k} + \sum_n \left(g^{(n)\,+}_{jk} A_{j'k'}^{j+n,k+n} + g^{(n)\,-}_{jk} A_{j'k'}^{j-n,k-n} \right)e^{-\tilde{\gamma}_{j'k'}t}\delta_{j'-k',j-k} ,\nonumber
\eea\end{widetext}
where the first line comes from performing the time derivative and the second line comes from direct substituion into the right hand side of Eq.~\eqref{coupledequations}. Making use of the linear independence of the functions $e^{-\tilde{\gamma}_{j'k'}t}$ to establish equality for each term, we rearrange to yield an expression 
%
\begin{equation}
\label{betterarg3}
\begin{split}
 A_{j'k'}^{(jk)} &(\tilde{\gamma}_{j'k'} - \tilde{\gamma}_{jk}) +\\  
 &\sum_n \left(g^{(n)\,+}_{jk} A_{j'k'}^{j+n,k+n} + g^{(n)\,-}_{jk} A_{j'k'}^{j-n,k-n} \right) = 0. 
 \end{split}
 \end{equation}
%
Since the $g^{(n)\,\pm}_{jk}$ are non-zero positive numbers, we immediately see that including an additional $n$-photon loss or gain channel must modify the coefficient $ A_{j'k'}^{(jk)}$. A corollary of this is that each coefficient $ A_{j'k'}^{(jk)}$ in a stripe is a function of \emph{every} $\tilde{\gamma}_{jk}$ and $g^{(n)\,\pm}_{jk}$ within that same stripe. Now, consider if a single $ A_{j'k'}^{(jk)} $ were zero. This implies that $\sum\limits_n g^{(n)\,+}_{jk} A_{j'k'}^{j+n,k+n} + g^{(n)\,-}_{jk} A_{j'k'}^{j-n,k-n} =0$. This requires an incredible amount of fine-tuning, since each $A_{j'k'}^{j\pm n,k\pm n}$ depends on all system parameters. Even if this holds, for the diagonal elements of the density matrix, the coefficients $A_{j'k'}^{j\pm n,k\pm n}$ are necessarily real and positive and so this condition can never be met. Thus, every coefficient $ A_{j'k'}^{(jk)} $ in the sum in Eq.~\eqref{densityDiscreteFourierRepeat} is non-zero.

Having shown that the Fourier coefficients $R_{j'k'}^{\nu\mu s}$ of the expectation value are generally non-zero, we see that any expectation value of the form $\expect{\hadmu \ha^\nu}_s (t)$ inherits a set of frequencies $\omega_j$ present in the $(\mu-\nu)$th stripe of the density matrix, and that this set of frequencies is unchanged by open system effects; while each frequency may contribute differently, it is still present in the signal for all finite time. This set is also unchanged by operator ordering, as it must be, since different orderings correspond to the same operator physically. Mathematically, this is because the commutation relation $[\ha,\had]=1$ preserves the difference in powers of the creation and annihilation operators; powers of both are removed together. Because all frequencies that are present in a closed system expectation value are also present in the open system expectation values, any periodicity of the closed system expectation values must also be inherited by the open system expectation values (The effect of the real part of $\tilde{\gamma}_{jk} $ coming from open system effects is to rescale the Fourier coefficients). Thus, we have re-derived our results: recurrences of the closed system are inherited by the open system irrespective of operator ordering.

We now turn to our second task, deriving analytic solutions to the Kerr system's evolution at zero temperature with single-photon loss utilizing this Fock-space representation in terms of diagonal stripes of the density matrix. We begin by truncating the Fock space at some $N\gg \expect{\hat{n}}$, such that for each terminating row and column of the density matrix
\bea\label{coupledequationsterminating}
\frac{d}{dt} \rho_{Nk}(t) &=& - \tilde{\gamma}_{Nk} \rho_{Nk}(t) \quad \forall k,\nonumber \\
\frac{d}{dt} \rho_{jN}(t) &=& - \tilde{\gamma}_{jN} \rho_{jN}(t)\quad \forall j.
\eea 
This allows the final density matrix elements of each diagonal stripe to be determined exactly
\bea\label{coupledequationstermsol}
\rho_{Nk}(t) &=& \rho_{Nk}(0)\,e^{- \tilde{\gamma}_{Nk} t} \quad \forall k\nonumber \\
\rho_{jN}(t) &=&  \rho_{jN}(0) \,e^{- \tilde{\gamma}_{jN} t} \quad\forall j.
\eea
For simplicity, we focus on the bottom row of the density matrix, whose elements are given exactly by the second line of Eq.~\eqref{coupledequationstermsol}. From these, we will be able to recursively generate solutions to other density matrix elements within the same stripe, and generate the rest of the density matrix by complex conjugation. 

To do this, we substitute our ansatz solution in terms of the generalized discrete Fourier transform Eq.~\eqref{densityDiscreteFourierRepeat} into the simplified master equation for the Kerr system at zero temperature
\bea\label{coupledequationsSimp}
\frac{d}{dt} \rho_{jk}(t) = - \tilde{\gamma}_{jk} \rho_{jk}(t) + g^{(1)\,+}_{jk} \rho_{j+1,k+1}(t) ,
\eea 
yielding
\begin{widetext}
\bea\label{InsertAnsatz2}
\sum_{j',k'} \left(\left(\tilde{\gamma}_{j'k'}-\tilde{\gamma}_{jk}\right)A_{j'k'}^{(jk)} - g^{(1)\,+}_{jk}  A_{j'k'}^{(j+1,k+1)}\right)e^{-\tilde{\gamma}_{j'k'} t} \delta_{j'-k' - (j-k)} = 0.
\eea \end{widetext}

Here we proceed using the same methodology as we did in Eqs.~\eqref{betterarg2} and \eqref{betterarg3}: rearranging and isolating each term in the sum.

Since the frequencies within any individual stripe are distinct, each term in the sum is linearly independent and we conclude  
\bea\label{InsertAnsatz3}
A_{j'k'}^{(j,k)} = \frac{g^{(1)\,+}_{jk}}{\tilde{\gamma}_{j'k'}-\tilde{\gamma}_{jk}}A_{j'k'}^{(j+1,k+1)}.
\eea 
We now have a way to generate the Fourier coefficients in terms of the next coefficient in a stripe. From Eq.~\eqref{coupledequationstermsol}, we see that there is only a single nonzero term in the Fourier expansion of the bottom row of the density matrix. The coefficients are 
\bea\label{terminatingCoefficients}
A_{j'k'}^{(N,k)} = \begin{cases}
    \rho_{Nk}(0), & j'=N, k'=k\\
    0, & {\rm otherwise}
\end{cases}.
\eea 
Furthermore, because of the unidirectionality of the coupling in Eq.~\eqref{coupledequationsSimp} for zero temperature (that is, there is no way for population to travel from lower Fock states to higher ones), we find 
\bea\label{higherCoefficients}
A_{j'k'}^{(j,k)} = 0, & j>j',\,k>k'.
\eea

We now shift notation slightly and explicitly limit our attention to just the bottom left triangle of the truncated matrix: let $j\rightarrow N-i$ and $k \rightarrow N-m-i$, where $N$ is the Fock space truncation defined previously, and $m$ is how far off diagonal the stripe is, ranging from $0$ (the diagonal) to $N$ (the bottom left stripe consisting of a single element of the truncated density matrix), and $ i$ is the position within the stripe from its bottom-rightmost terminating element, ranging from $0$ to $N-m$. In this notation and making use of the truncated Fock space, we can now write an analytic solution for the density matrix of the open quantum system.

We begin by rewriting, in this new notation, our formal solution Eq.~\eqref{densityDiscreteFourierRepeat} 

\bea\label{densityDiscreteFourierSimplified}
\rho_{N-i,N-m-i}(t) = \sum\limits_{i'=0}^{i} A_{N-i',N-m-i'}^{(N-i,N-m-i)} e^{-\tilde{\gamma}_{N-i',N-m-i'}t},  \nonumber \\
\eea where we have made use of the result in Eq.~\eqref{higherCoefficients} that $A_{N-i',N-m-i'}^{(N,N-m)} = 0$ for $i'>i$ to truncate the sum at $i$ instead of $N-m$. We also rewrite in the new notation the recursion relation Eq.~\eqref{InsertAnsatz3}
\begin{widetext}
    \bea\label{RecursionRelation1}
\!\!A_{N-i',N-m-i'}^{(N-i,N-m-i)} =\frac{g^{(1)\,+}_{N-i,N-m-i}}{\tilde{\gamma}_{N-i',N-m-i'}-\tilde{\gamma}_{N-i,N-m-i}}A_{N-i',N-m-i'}^{(N-i+1,N-m-i+1)}\nonumber \\
=\left(\prod\limits_{u=1}^{i-1}\frac{g^{(1)\,+}_{N-i+u,N-m-i+u}}{\tilde{\gamma}_{N-i',N-m-i'}-\tilde{\gamma}_{N-i+u,N-m-i+u}}\right)A_{N-i',N-m-i'}^{(N,N-m)}\,\,,
\eea 
\end{widetext}
where in the second line we have applied the recursion relation $i-1$ times and have assumed $i>0$ and $i'\leq i$ (again, because of the result in Eq.~\eqref{higherCoefficients}).

Defining the function
\bea\label{fNmi}
f_{Nmii'} = \begin{cases}
    \prod\limits_{u=1}^{i-1}\frac{g^{(1)\,+}_{N-i+u,N-m-i+u}}{\tilde{\gamma}_{N-i',N-m-i'}-\tilde{\gamma}_{N-i+u,N-m-i+u}} & i>0, \\
    1 & i=0, \\
\end{cases}
    \nonumber \\
\eea and substituting Eq.~\eqref{RecursionRelation1} into our formal solution Eq.~\eqref{densityDiscreteFourierSimplified}, we find
\begin{equation}
\label{denssol}
\begin{split}
&\rho_{N-i,N-m-i} (t)\\ 
&\qquad =  \sum\limits_{i'=0}^{i} f_{Nmii'}
A_{N-i',N-m-i'}^{(N,N-m)}e^{-\tilde{\gamma}_{N-i',N-m-i'} t}.  
\end{split}
\end{equation}

We have reduced the problem of the dynamics of a single stripe of the density matrix elements $\rho_{N-i,N-m-i} (t)$ to finding the matrix of coefficients $A_{N-i',N-m-i'}^{(N,N-m)}$, which we can now derive from boundary terms, and evaluating Eq.~\eqref{denssol}. Notably, now the coefficients $A_{N-i',N-m-i'}^{(N,N-m)}$ are independent of $i$ (the location within the stripe) and can form a matrix, which is spanned by the coordinates $i'$ and $m$. One can then recursively solve for the elements $A_{N-i',N-m-i'}^{(N,N-m)}$ for each $i$ at $t=0$ and making use of the boundary condition of the known initial state 
\bea \label{BoundaryCondition}
\rho_{N-i,N-m-i} (0) =  \sum\limits_{i'=0}^{i} f_{Nmii'}
A_{N-i',N-m-i'}^{(N,N-m)}. 
\eea 

For $i=0$ (the terminating row of the density matrix) we already have the solution for density matrix element evolution in Eq.~\eqref{coupledequationstermsol} and have written the values of the coefficients in Eq.~\eqref{terminatingCoefficients}. Since there is a single term only, we conclude for the $i=0$ case $A_{N,N-m'}^{(N,N-m)} = \rho_{N,N-m}(0)$. By then considering the $i=1$ case and using the boundary condition (Eq.~\eqref{BoundaryCondition}), we can recursively generate the coefficients for larger and larger $i$ up to its maximum value of $N-m$. For $i=1$, we find
\bea\label{cases2}
\rho_{N-1,N-m-1} (0) &=& f_{Nm10} A_{N,N-m}^{(N,N-m)} \nonumber \\
&& + f_{Nm11} A_{N-1,N-m-1}^{(N,N-m)}\nonumber \\
\rightarrow A_{N-1,N-m-1}^{(N,N-m)} &=& \frac{\rho_{N-1,N-m-1} (0) - f_{Nm10} A_{N,N-m}^{(N,N-m)}}{f_{Nm11} } \nonumber \\
\eea 
where in the second line we have made use of the result from $i=0$. For the general case we find 
\begin{widetext}\bea\label{cases3}
A_{N-i,N-m-i}^{(N,N-m)} &=& \begin{cases}
\rho_{N,N-m}(0) & i=0, \\
\frac{1}{f_{Nmii}} \left( \rho_{N-i,N-m-i}(0) - \sum\limits_{i'=0}^{i-1} f_{Nmii'} A_{N-i',N-m-i'}^{(N,N-m)} \right)& i>0.
\end{cases}
\eea\end{widetext} 
The coefficients generated from Eq.~\eqref{cases3} can then be substituted into Eq.~\eqref{denssol} to give the analytic solution for the full time evolution of the density matrix elements. 


This recipe scales poorly ($\sim \alpha_0^6$), is tedious to derive, fails numerically for the closed system ($f_{Nmii'}$ develops singularities), and is memory-intensive. However, it has the nice property that, since it front-loads the difficulty of the problem into calculating the matrix of coefficients $A$, it is intrinsically stable in time. With this recipe, we have generated a movie giving the qualitative features of the evolution of the bosonic anharmonic oscillator for a small value of $\alpha_0$ \cite{WignerVideo}.

Analytic solutions to non-linearly evolving open quantum system dynamics have thus far been rare to find in quantum optics. It is our hope that this method, though cumbersome, may inspire others to find more solutions. 

\section{Kitten State Expectation Values}\label{app:expectkittens}

In this section, we derive the three cases in Eq.~\eqref{kittenexpectexpanded} explicitly, solidifying the relationship between the $N$-kitten states and the operators $\hat{X}_{\theta}^n$ via their shared symmetry $\mathbb{Z}_n$ (the $n$-fold cyclic group symmetry). 

Recall from Eq.~\eqref{XThetaDecomp} that an operator $\hat{X}_{\theta}^n$ is decomposable into a sum of symmetrically ordered operators $\{\hadmu \ha^\nu\}_{\rm sym}$, where the difference $|\mu-\nu|\leq n$ for each term. By the results of Appendix \ref{app:reccurencethms}, recurrence times are independent of operator ordering and open system effects, so it will suffice to study normally ordered operators $\hadmu \ha^\nu$ of the closed quantum system.

The question now is, given an $N$-kitten state $\ket{\psi(t_{M,N})}$ as defined in Eq.~\eqref{KerrState}, what is $\langle\hadmu \ha^\nu\rangle$ and when is the semiclassical TWA a reasonable approximation to it?  If, for some $\mu$ and $\nu$, this expectation is independent of $M$ and $N$, then that expectation value must be entirely calculable in terms of semiclassical behavior since it does not depend on the quantum structure of the state (i.e.\ which kitten state has formed). 
To answer this question, it is sufficient to consider the following three cases: a) $\mu=\nu$, b) $\mu - \nu = p N$, and c) $\mu - \nu \neq p N$ where $p$ is a nonzero integer.

\vspace{1em}

{\textbf{Case a)}} $\mu=\nu$: Since $[\hat{H}_{\rm Kerr},\hat{n}]=0$, all operators proportional to powers of the number operator are conserved in the evolution $\ket{\psi (t_{M,N})} = e^{i \hat{H}_{\rm Kerr} t_{M,N}} \ket{\psi (0)}$. Thus we conclude $\langle\hadmu \ha^\nu\rangle=|\alpha_0|^{2\mu}$ when $\mu=\nu$. This is an entirely semiclassical quantity (and easy to calculate, as it does not even require the semiclassical approximation of the unitary dynamics).

\vspace{1em}
{\textbf{Case b)}} $\mu - \nu = p N$: 
First note that $\ket{\psi(t_{M,N})}$ is an eigenstate of $\ha^N$ with eigenvalue $\pm\alpha_0^N$, where the sign is negative only if $N$ is even and $M$ is odd. 
To see this, consider applying $\ha^N$ to a single term in the superposition. Each application of $\ha$ multiplies the coherent state 
by the complex number: $\alpha_0 e^{\frac{i\pi(2k+M)}{N}}$ if $N$ is even and $\alpha_0 e^{\frac{i\pi 2k}{N}}$ if $N$ is odd. After $N$ applications the amplitude of this number is independent of $k$ with the sign dependence noted above.

Thus, if the difference of $\mu$ and $\nu$ is an integer multiple of $N$, we can utilize the eigenproperty to pull out $\mu-\nu$ powers of $\alpha_0^*$ and are left with an expectation value of a power of the number operator as in the first case. Explicitly,

\begin{equation}\label{c1}
    \langle \hat{a}^{\dag \mu} \ha^{\nu}\rangle= (\pm 1)^p \alpha_0^{*\mu-\nu}
    |\alpha_0|^{2\nu} \qquad \mu-\nu=p\,N,
\end{equation}
where, again, the sign is negative if $N$ is even and $M$ is odd and positive otherwise. 

This quantity is $M$ and $N$ dependent and thus will not be captured by semiclassical calculations that are insensitive to the particular quantum state. Hence, the TWA will not give the correct value here. 

\vspace{1em}
{\textbf{Case c)}} $\mu - \nu \neq pN$: 
First we show the following. For $2\pi|\alpha_0| \gg N$, the expectation value $\bra{\psi(t_{M,N})} \ha^q\ket{\psi(t_{M,N})}$ tends to zero for $0<q<N$ and is exactly zero for all $N$ in the infinite-$|\alpha_0|$ limit. Here the integer $q$ plays the role of $pN$ (which must be an integer since $\mu-\nu$ is an integer). For now we assume $p<1$ so that $q<N$, but this restriction will be relaxed later.

The expectation value $\bra{\psi(t_{M,N})} \ha^q\ket{\psi(t_{M,N})}$ is rewritten in terms of an analytic function $f(\omega)$, such that 
\bea\label{defexpectfomega}
\bra{\psi(t_{M,N})} \ha^q \ket{\psi(t_{M,N})} = \tilde{\alpha}_0^q  f(\omega_N^q) + \mathcal{O}(\braket{\alpha_i}{\alpha_j}) ,
\eea 
where $\omega_N$ is an $N$th root of unity $e^{2\pi i /N}$ and we have defined $\tilde{\alpha}_0 = e^{\frac{i\pi M}{N}} \alpha_0$ for $N$-even $M$-odd and $\tilde{\alpha}_0=\alpha_0$ otherwise. $\mathcal{O}(\braket{\alpha_i}{\alpha_j})$ goes to zero in the limit of orthogonal coherent states (that is, $2\pi|\alpha_0| \gg N$ for coherent states distributed uniformly around the circle with radius $|\alpha_0|$). $f(\omega)$ is the sum of a finite geometric sequence
\bea\label{fdef}
f(\omega) = 1+\omega+\omega^2+\omega^3+\dots + \omega^{N-1}=\frac{1-\omega^N}{1-\omega}.
\eea 

 \begin{figure}[t] 
	\includegraphics[width=\linewidth]{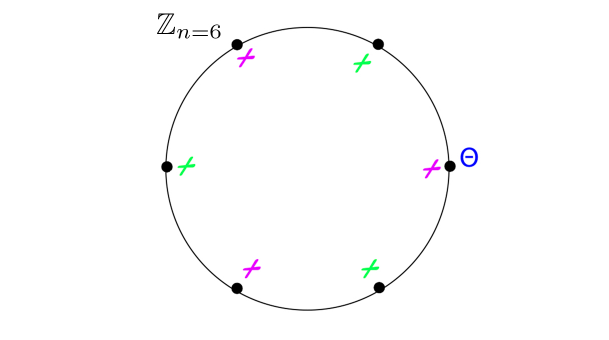} 
	\caption[]{The cyclic group $\mathbb{Z}_n$ for $n=6$. Initially, all states have the same phase, marked by the position on the circle by a blue $\theta$ (as all the phases are measured with respect to this un-moving point). After a single application of $\ha$, they will have phases equally distributed over the circle at the location of the black dots. After a second application of $\ha$, they will now be equally distributed around the three pink markers. After a third application, they will be equally distributed around two points: $\theta$ and its antipode. After a fourth application, they will be equally distributed around the three green markers. And after a fifth application, they will be equally distributed around the black dots again before finally ``recohering'' after a sixth ($n$th) application at the dot with the blue theta. At each stage the group action maintains an equal distribution over the support: the roots of unity corresponding to each way of factoring $n$.
 
For a video describing this, see \cite{6kittenVideo}.}
	\label{Zn}
\end{figure}

Since $\omega^{qN}=1^q=1$, it follows that $f(\omega_N^q)=0$. Hence, the corresponding term in the expectation value in Eq.~\eqref{defexpectfomega} is zero as well.

This result can also be understood geometrically, considering the unit circle and the cyclic group $\mathbb{Z}_N$ as illustrated in Fig.~\ref{Zn} (as well as in the accompanying video \cite{6kittenVideo}); at each stage of the group action (rotating the phase of each state vector by the phase of its coherent state), an initially equal distribution over the support remains equally distributed, even as the structure of the support changes due to common factors within the roots of unity.

Both this property and the eigenvalue relation above will also apply to the associated mixed states distributed uniformly in the same manner. The only difference is that now this property holds exactly even outside the large-$\alpha_0$ limit. Since the density matrix is diagonal, there are no contributions from terms $\mathcal{O}(\braket{\alpha_i}{\alpha_j})$
as there are for the superposition. 

We now know the expectation value in Eq.~\eqref{defexpectfomega} is approximately zero for $q<N$. From this, we can conclude that the expectation value $\expect{\hadmu \ha^\nu}$ where $\mu - \nu = pN$ with $p \notin \mathbb{Z}$ is also approximately zero using the eigenstate property. Taking $\mu<\nu$, we rewrite the expectation value as $\expect{\hadmu \ha^{\nu-\mu} \ha^{\mu}}$. The action of $\hadmu$ to the left and $\ha^{\mu}$ to the right result in matching rotations of the phases in the kitten state and an overall multiplication of the state by $|\alpha_0|^{2\mu}$. Thus, the expectation value reduces to $|\alpha_0|^{2\mu}\expect{ \ha^{\nu-\mu}}$, and we can make use of our results from Eq.~\eqref{defexpectfomega} with $q=\nu-\mu$ to say that this zero, provided $\nu-\mu < N$. If instead $\nu-\mu> N$, we make use of the eigenstate property to pull out $N$ powers of the annihilation operator $\ha$ and replace them with $\tilde{\alpha}_0^N$, repeating the process $w$ times until there are $q=\nu-\mu-wN<N$ powers of $\ha$ remaining in the expectation value and a pre-factor $\tilde{\alpha}_0^{wN}$. Since $\nu-\mu$ is not an integer multiple of $N$, $q=\nu-\mu-wN$ is not an integer multiple of $N$, and the expectation value is approximately zero. Similarly, for $\mu>\nu$, taking complex conjugation to give the associated expressions. 

Thus, for a sufficiently large $|\alpha_0|$, the expectation value is approximately zero and can be calculated semiclassically. Note that when the number of kittens $N$ becomes large enough such that $ 2\pi |\alpha_0| \lesssim N$, their non-orthogonality contributes to oscillations, giving rise to the non-precision of the recurrences (due to neighboring high-$N$ kitten states) as well as the deviations from the TWA at short times (also high-$N$ kitten states) in Fig.~\ref{AverageOscillationTime}. 

To summarize, we have rederived the three cases of Eq.~\eqref{kittenexpectexpanded}. We know that the recurrences for normally ordered expectation values given in Eq.~\eqref{period} that we calculated in the Heisenberg picture is a manifestation of the symmetry of kitten states being that of the symmetric group $\mathbb{Z}_n$, which is shared by operators $\hadmu \ha^\mu$ when $\mu-\nu =p N$, with $p\in \mathbb{Z}^\pm$. By the various results of the previous sections, we can also appreciate that this is true of symmetric products of powers of $\had$ and $\ha$. The operators $\hat{X}_\theta^n$, when expanded in such products, will contain terms with $\mu-\nu=\pm n$, hence the relationships between the operators $\hat{X}_\theta^n$ and the $N$-kitten states when $n=pN$.

\section{Example of a Trivial Quantum-to-Classical Transition: a Freely Decaying Cat State}\label{app:DecayCat}

At short times, the freely-decaying cat state provides a simple example of a trivial quantum-to-classical transition---that is, a transition where the quantum state is drowned out by classical noise. As seen in Eqs.\,(\ref{decaykittengen}--\ref{CatWigner}), the Wigner function for a freely-decaying cat state has the form at short times 

\begin{widetext}\begin{equation}\label{decaycateq}\begin{split}
W(X,P,t) &= 
\frac{1}{2}W_0(X,P-X_0 e^{-\frac{\gamma t}{2}}) + \frac{1}{2}W_0(X,P+X_0 e^{-\frac{\gamma t}{2}})+ \braket{+X_0}{-X_0}^{1-e^{-\gamma t}} \sin(2X X_0)W_0(X,P) \\
&\approx \frac{1}{2}W_0(X,P-X_0) + \frac{1}{2}W_0(X,P+X_0)+ e^{-X_0^2 \gamma t}\sin(2X X_0)W_0(X,P),
\end{split}
\end{equation} where in the second-line we've taken the short-time limit $\gamma t \ll 1$. Defining

\bea\label{closedcat}
W_{\rm closed\,cat}(X,P) &=&\frac{1}{2}W_0(X,P-X_0) + \frac{1}{2}W_0(X,P+X_0)+\sin(2X X_0)W_0(X,P),\nonumber \\
W_{\rm mix\,cat}(X,P) &=&\frac{1}{2}W_0(X,P-X_0) + \frac{1}{2}W_0(X,P+X_0),
\eea and $p(t) = e^{-X_0^2 \gamma t}$, we see that (\ref{decaycateq}) has the form of (\ref{WConvex}), that is,

\begin{equation}\label{WConvex2}\begin{split}
W_{\rm open \,cat}(X,P,t) =  p(t) W_{\rm closed\,cat}(X,P) + (1-p(t)) W_{\rm mix\,cat}(X,P).
\end{split}
\end{equation}\end{widetext}

Considering the difference defined in (\ref{DifferenceConvex}) for an observable, for instance, $\hat{X}^n\hat{P}^m$, we obtain

\begin{equation}\label{differencecat}
\delta\langle \hat{X}^n\hat{P}^m \rangle = e^{-X_0^2 \gamma t} \int dX dP  X^n P^m \sin(2X X_0)W_0(X,P).
\end{equation} Note that the pre-factor $e^{-X_0^2 \gamma t}$ is independent of $n$ and $m$, and is the only time-dependence in the expression; if we applied the test-implemented in Fig. \ref{convexitytest}, we would find that the measurement statistics are perfectly consistent with a trivial quantum-to-classical transition. 

The statements made above hold true for the long-time behavior as well as for higher-order kitten states with minor modifications. For long-time behavior, the form of $p(t)$ is changed and, more interestingly, the Wigner functions become time-dependent to include the effects of energy loss (but not decoherence)

\begin{widetext}\begin{equation}\label{energyloss}
\begin{split}
W_{\rm closed\,cat}&(X,P,t) =\frac{1}{2}W_0(X,P-X_0e^{\frac{-\gamma t}{2}}) + \frac{1}{2}W_0(X,P+X_0e^{\frac{-\gamma t}{2}})+\sin(2X X_0 e^{\frac{-\gamma t}{2}})W_0(X,P),\nonumber\\
W_{\rm mix\,cat}&(X,P,t) =\frac{1}{2}W_0(X,P-X_0e^{\frac{-\gamma t}{2}}) + \frac{1}{2}W_0(X,P+X_0e^{\frac{-\gamma t}{2}}).
\end{split}
\end{equation}\end{widetext}

 For higher-order kitten states where there are more than three kittens, one must introduce multiple classical mixed states, each with a distinct $p(t)$. This is necessary to account for the different rates at which the pair-wise superpositions decohere, which increases with their seperation. This gives rise to a more general noise model where the open quantum system's state is a convex combination of the closed quantum system's state and several classical states: $W_{\rm open}(\alpha,t) = p_0(t) W_{\rm closed}(\alpha,t) + \sum_i p_i(t) W_{\rm i}(\alpha,t)$ with $\sum_i p_i(t) = 1$. For the decaying cat state, the $W_{\rm i}(\alpha,t)$ correspond to classical mixtures of coherent states with equal separation for $i> 0$. Since here there is no single classical background, the test used in Fig. \ref{convexitytest} will not be useful. However, the measurement statistics can still be used to detect whether a trivial quantum-to-classical transition is present as, again, the same time-dependent factors $p_i(t)$ are inherited by all observables independently of expectation value order. 

Lastly, we note that while the difference between the closed system and the classical state expectation values all decay at the same rate, this is not true of the raw expectation values themselves. To see this, consider, e.g., the open-system evolution of the expectation value $\langle\hat{X}^n\rangle$ in the adjoint master equation

\bea\label{adjointXn}
\frac{d}{dt} \langle \hat{X}^n \rangle &=& - \frac{\gamma}{2} \left(\langle\had [\ha,\hat{X}^n]\rangle +  [\hat{X}^n,\had]\had\rangle\right) \nonumber \\
&=& \frac{-n \gamma}{2} \langle \hat{X}^n \rangle + \frac{n(n-1)}{4} \langle \hat{X}^{n-2} \rangle. 
\eea Here, we see higher-order expectation values will decay faster in their value, regardless of the initial state. Nonetheless, this has nothing to do with decoherence and everything to do with energy loss. When we look at the difference between expectation values of the closed system and of the classical system, as we have done in  (\ref{differencecat}), we see that all expectation values approach classical behavior at the same rate, which is a tell-tale sign of a trivial quantum-to-classical transition.

\end{document}